\documentclass[fleqn,usenatbib]{mnras}

\usepackage[T1]{fontenc}

\DeclareRobustCommand{\VAN}[3]{#2}
\let\VANthebibliography\thebibliography
\def\thebibliography{\DeclareRobustCommand{\VAN}[3]{##3}\VANthebibliography}

\usepackage{graphicx}	% Including figure files
\usepackage{amsmath}	% Advanced maths commands
\usepackage{amssymb}	% Extra maths symbols
\usepackage[utf8]{inputenc}
\usepackage{textcomp}
\usepackage[varg]{txfonts}
\usepackage[version=3]{mhchem}
\usepackage{upgreek}
\usepackage[outdir=plot/]{epstopdf}
\usepackage{multirow}

\newcommand\tablefoot[1]{%
  \par\vspace{2ex}
  \noindent
  \begin{minipage}{\linewidth}
    {\small\bfseries Notes.}~%
    \small
    \ignorespaces
    #1%
  \end{minipage}%
}
\newcommand*\tablefootmark[1]{%
  \unskip
  \hbox{\textsuperscript{\normalfont\ignorespaces#1}}%
  \,%
  \ignorespaces
}
\newcommand\tablefoottext[2]{%
  \hbox{\textsuperscript{\normalfont({\ignorespaces#1})}}%
  ~%
  \ignorespaces
  #2\ \ignorespaces%
}

\title[MmW spectrum of 2-PA]{Millimeter-wave spectrum of 2-propanimine}

\author[L. Zou et al.]{
Luyao Zou,$^{1}$\thanks{Current address: Laboratoire de Physico-Chimie de l'Atmosph\`ere, 
Universit\'e du Littoral C\^ote d'Opale, 189 A Ave.\ Maurice Schumann, F-59140 Dunkerque, France. E-mail: luyao.zou@univ-littoral.fr} %
Jean-Claude Guillemin,$^{2}$ %
Arnaud Belloche,$^{3}$ %
Jes K.\ J\o{}rgensen,$^{4}$\newauthor %
Laurent Margul{\`e}s,$^{1}$ %
Roman A.\ Motiyenko$^{1}$ %
and Peter Groner$^{5}$ %
\\
$^{1}$Univ.\ Lille, CNRS, UMR 8523 -- PhLAM -- Physique des Lasers Atomes et Mol\'ecules, F-59000 Lille, France\\
$^{2}$Univ Rennes, Ecole Nationale Sup\'erieure de Chimie de Rennes, CNRS, ISCR-UMR 6226, F-35000 Rennes, France\\
$^{3}$Max-Planck-Institut f\"ur Radioastronomie, Auf dem H\"ugel 69, 53121 Bonn, Germany\\
$^{4}$Niels Bohr Institute, University of Copenhagen, Copenhagen, Denmark\\
$^{5}$Department of Chemistry, University of Missouri -- Kansas City, 5100 Rockhill Rd., Kansas City, MO 64110, USA
}

\date{Accepted XXX. Received YYY; in original form ZZZ}

\pubyear{2022}

\begin{document}
\label{firstpage}
\pagerange{\pageref{firstpage}--\pageref{lastpage}}
\maketitle

\begin{abstract}
Up to date, only 6 imines have been detected in the interstellar medium. 
The 3-carbon imine, 2-propanimine (\ce{(CH3)2C=NH}), is predicted to be the structural isomer with the lowest energy in the \ce{C3H7N} group, 
and appears to be a good candidate for astronomical searches.
Unexpectedly, no microwave or millimeter wave spectrum is available for 2-propanimine.
In this work, we provide the first high resolution millimeter wave spectrum of 2-propanimine and its analysis. 
With the guide of this laboratory measurement, we aim to search for 2-propanimine in 
two molecule-rich sources Sgr~B2(N) and IRAS~16293-2422 using observations 
from the Atacama Large Millimeter/submillimeter Array (ALMA).
Starting from a synthesized sample, we measured the spectrum of 2-propanimine from 50 to 500~GHz, 
and the ground state lines are successfully assigned and fitted using XIAM and ERHAM programs 
with the aid of theoretical calculations. 
The barriers to internal rotation of the two \ce{CH3} tops 
are determined to be 531.956(64)~cm$^{-1}$ and 465.013(26)~cm$^{-1}$ by XIAM. 
These data are able to provide reliable prediction of transition frequencies for astronomical search. 
Although a few line matches exist, 
no confirmed detection of 2-propanimine has been found in the hot molecular core Sgr~B2(N1S) and the Class 0 protostar IRAS~16293B.
Upper-limits of its column density have been derived, 
and indicate that 2-propanimine is at least 18 times less abundant than methanimine in Sgr~B2(N1S), 
and is at most 50--83~\% of methanimine in IRAS~16293B.
\end{abstract}


%%%%%%%%%%%%%%%%%%%%%%%%%%%%%%%%%%%%%%%%%%%%%%%%%%

%%%%%%%%%%%%%%%%% BODY OF PAPER %%%%%%%%%%%%%%%%%%

\section{Introduction}

In the interstellar medium (ISM), N-bearing compounds are important prebiotic molecules because of their link to the chemical synthesis of amino acids, the building blocks of proteins \citep{Holtom2005ApJ}.
The chemistry of aliphatic amines\footnote{Organic molecules carrying a \ce{C-N} single bond.} 
and imines\footnote{Organic molecules carrying a \ce{C=N} double bound.} is not well understood;
their detection in the ISM is also limited to only a few sources, 
in great contrast to their O-bearing counterparts alcohols\footnote{Organic molecules carrying a \ce{O-H} single bond and this C is not directly connected to an aromatic ring}, 
aldehydes\footnote{Organic molecules carrying a \ce{-CHO} terminal, where C and O are connected with double bond.}, 
and ketones\footnote{Organic molecules carrying a \ce{C=O} double bond and this C is connected to two other C atoms.}.
Up to date, only six imines (including geometric isomers) have been detected. 
The simplest methanimine (\ce{CH2NH}) is observed in various types of sources \citep{Godfrey1973ApJL, WidicusWeaver2017ApJS}.
In the group of two carbon imines, ($E$)- and ($Z$)- ethanimine (\ce{CH3CH=NH}) \citep{Loomis2013ApJL}, 
and ($E$)- and ($Z$)- cyanomethanimine (\ce{NH=CHCN}) \citep{Zaleski2013ApJ}, have been observed in the hot core Sgr~B2(N).
One 3-carbon imine, propargylimine (\ce{HCCCH=NH}), has been detected in the quiescent cloud G+0.693-0.027 \citep{Bizzocchi2020AA}.
All observed imines belong to aldimines\footnote{Imines where the C in the \ce{C=N} bond is connected to an H atom and a C atom}, 
which are analogous to the structure of aldehydes.
No ketimine\footnote{Imines where the C in the \ce{C=N} bond is connected to two other C atoms}, 
with a structure analogous to that of ketone, has been discovered in space yet. 
For amines, the simplest amine, methylamine (\ce{CH3NH2}), has been observed in only four sources: Sgr~B2(N) \citep{Kaifu1974ApJL,Fourikis1974ApJL}, G10.47+0.03 \citep{Ohishi2019PASJ}, NGC 6334F \citep{Ohishi2019PASJ}, and NGC 6334I \citep{Bogelund2019AA-NGC6334}.
Aminoacetonitrile (\ce{NH2CH2CN}) has been detected in Sgr~B2(N) \citep{Belloche2008AA}.
The other three larger amines, vinylamine (\ce{C2H3NH2}) \citep{Zeng2021ApJL}, ethylamine (\ce{C2H5NH2}) \citep{Zeng2021ApJL}, and ethanolamine (\ce{NH2CH2CH2OH}) \citep{Rivilla2021PNAS}, have been recently detected or tentatively detected in G+0.693-0.027. 

The difficulty to detect amines and imines in the ISM prevents us from further understanding their interstellar chemistry.
To explore the possibility of detecting more amines and imines, 
\citet{Sil2018ApJ} performed a survey of all isomeric groups of amines and imines of one to three carbons using theoretical calculation.
In the \ce{C3H7N} group, 2-propanimine, \ce{(CH3)2C=NH}, is the most stable structural isomer and is the simplest ketimine. 
Following the minimum energy principle \citep{Lattelais2010AA}, 2-propanimine is a good interstellar molecule candidate. 
\citet{Sil2018ApJ} claimed that (Z)-1-propanimine, which is 6.45~kcal/mol (equivalent to 3246~K) higher in energy than 2-propanimine, 
has a higher chance of detection because of its larger dipole moment value and higher estimated abundance from astrochemical modeling. 
A recent search for 1-propanimine, however, has reported nondetection in Sgr~B2(N) \citep{Margules22}.
That being said, it is unreasonable to rule out 2-propanimine as a potential interstellar molecule, 
considering the variety of physical and chemical environments in the ISM, and the uncertainties of theoretical calculation and modeling. 
2-propanimine possesses a structure analogous to acetone, with an \ce{N-H} moiety replacing the \ce{O} atom in acetone. 
Because acetone is a well-known interstellar molecule detected in multiple hot core regions \citep{Combes1987AA, Snyder2002ApJ, Friedel2005ApJL, Friedel2008ApJ, Friedel2012ApJS, Peng2013AA, Rolffs2011AA, Isokoski2013AA, Zou2017ApJ, Suzuki2018ApJS}, \
low-mass and intermediate mass protostars \citep{Jorgensen2011AA, Lykke2017AA, Fuente2014AA},
it is also interesting in the future to investigate the chemical similarity and difference between 2-propanimine and acetone in the ISM. 

From the spectroscopic point of view, 2-propanimine is a molecule in $C_s$ molecular symmetry
and is associated with two slightly nonequivalent \ce{CH3} internal rotors.
The inequality arises from the slightly tilted \ce{N-H} bond, which creates nonequivalent barriers to internal rotation of the two \ce{CH3} tops. 
In such case, the tunnelling effect of internal rotation splits each rotational level into 9 components, 
with one non-degenerate state belonging to the $A_1$ symmetry, and four doubly degenerate states belonging to the $E$ symmetry ($E_1$, $E_2$, $E_3$, and $E_4$). 
Each rotational line thus splits into five components, which can be labeled by two symmetry numbers $\sigma_1, \sigma_2$, or conventionally by symmetry labels $A$ and $E$. 
These components correspond to $\sigma_1, \sigma_2=(0,0)$, $(0,1)$, $(1,0)$, $(1,1)$ and $(1,2)$, 
or correspondingly, the $AA$, $AE$, $EA$, $EE$, and $EE'$ states. 
The ($1,1$) and ($1,2$) are in most cases degenerate, and they separate only at low $J$ and high $K_a$. 
The relative weight of the five components is 2:2:2:1:1. 
If we consider the spin statistics of the six H-atoms of the two \ce{CH3} tops, 
the total spin statistical weight is then 16:16:16:8:8. 
Because the barriers to internal rotation of the two \ce{CH3} tops of 2-propanimine are only slightly different, 
the tunnelling splittings caused by the two tops are nearly equal for the ground state. 
Therefore, we expect to see sets of almost evenly spaced quadruplets of equal intensity in the spectra, 
except when the degeneracy of the ($1,1$) and ($1,2$) states is lifted at low $J$ values to reveal the quintuplets. 

Under terrestrial conditions, 2-propanimine is unstable and is usually produced in pyrolysis \citep{ZhengS2003JCP} or flame reactions \citep{TianZY2009ProcCombIns}.
High resolution spectroscopic study of 2-propanimine is still lacking,  
and surprisingly, even no microwave studies of 2-propanimine are available to our knowledge. 
In this study, we present the first high resolution millimeter-submillimeter wave spectrum of 2-propanimine from a pure synthesized sample. 
The pure sample, which is free from the inevitable contamination of side products in pyrolysis and flame reactions, 
allows us to obtain high quality rotational spectrum of 2-propanimine between 50 and 500~GHz.  

\section{Experimental Methods}

The 2-propanimine was synthesized from the reaction of \ce{KOH} and precursor 2-amino-2-methylpropanenitrile. 
The synthesized sample was collected and preserved under dry ice temperature. 
Spectroscopic measurement was performed using the fast absorption spectrometer at Lille.
Throughout the measurement, the sample was submerged in an ethanol cold bath at $-60$~\textdegree{}C,  
and a minimum flow of the sample vapor was maintained between 8--20~$\upmu$Bar. 
Spectrum was measured between 50--110~GHz, 150--330~GHz, and 360--500~GHz using commercial frequency amplifier-multiplier chains 
and frequency modulation and second harmonics detection techniques. 
Quantum chemical calculation using Gaussian16 \citep{Gaussian16} was also performed to guide the spectral analysis 
by estimating the structure, rotational constants, dipole moments, 
and the barriers to internal rotation of the two \ce{CH3} groups of 2-propanimine. 
The experimental spectrum was processed using a custom spectral assignment software, 
and fitted using XIAM \citep{Hartwig1996ZNatA, Herbers2020JMS} and ERHAM \citep{Groner1997JCP,Groner2012JMS} programs
with an iterative approach starting from the predicted spectrum from quantum chemical calculation. 
More technical details can be found in Appendix~\ref{app:exp}.

\section{Results and Discussion}\label{sect:result}

\begin{figure}
  \centering
  \includegraphics[width=0.4\textwidth]{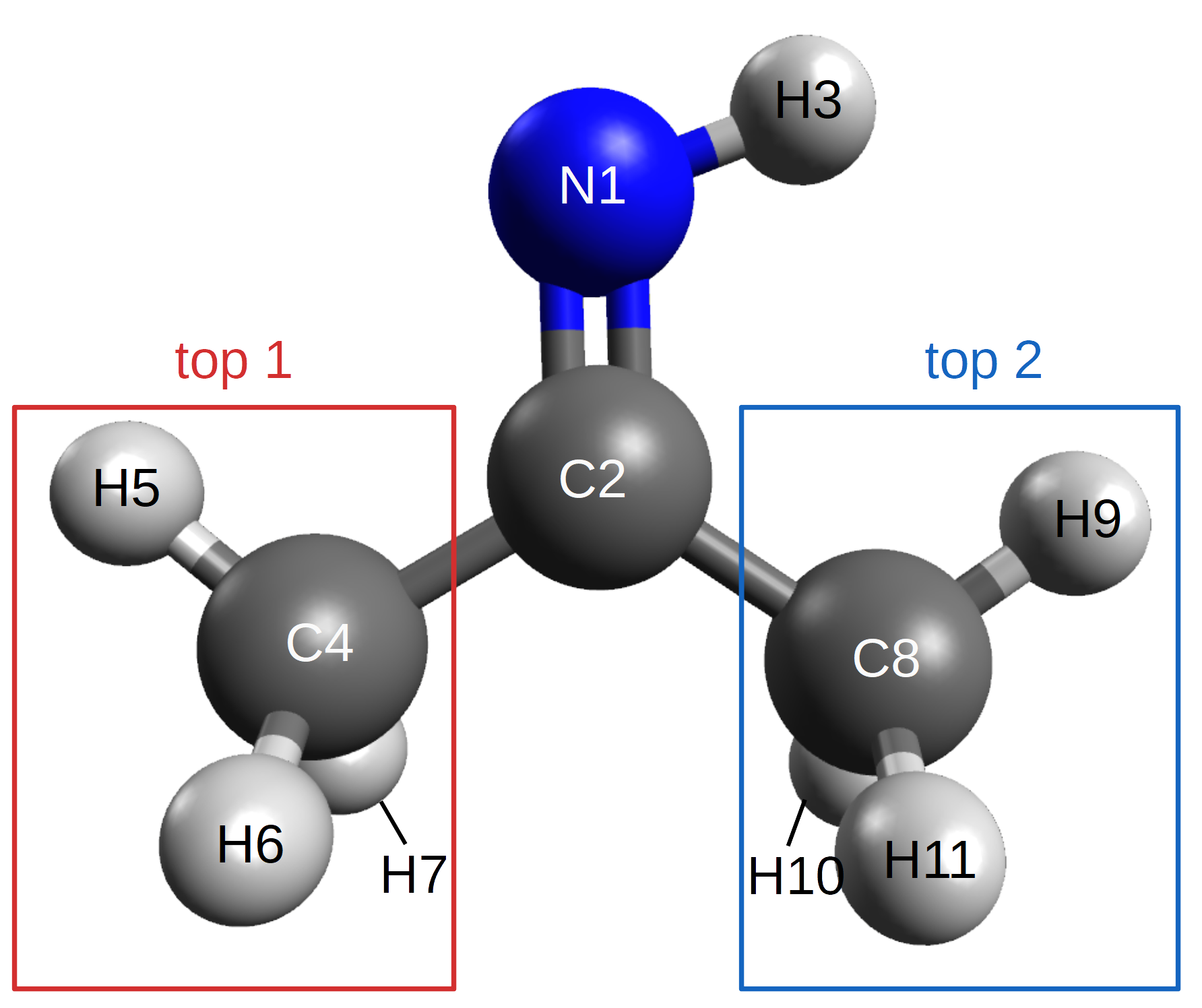}
  \caption{Optimized molecular structure of 2-propanimine.  ``Top 1'' refers to the \ce{CH3} group away from the \ce{N-H} bond, 
  and ``top 2'' refers to the one close to the \ce{N-H} bond.
  Atoms N1, C2, C4, C8, H3, H5, and H9 are all in the same plane, which is the symmetry plane of the molecule. 
  Atoms H6 and H7 are symmetric with respect to this plane, as are atoms H10 and H11.\label{fig:2PA-structure}}
\end{figure}

All quantum calculation methods produce similar equilibrium structures of 2-propanimine (see Appendix~\ref{app:calc-xyz}).
The potential energy surface (PES) scan with the \ce{N-H} bond twisting with respect to the \ce{C-C(N)-C} plane supports that 
the minimum energy structure is when the \ce{N-H} bond lies within the same plane of the carbon skeleton, leading to the $C_s$ symmetry of the molecule. 
Supplementary 2-dimentional PES scan (see Appendix~\ref{app:mol-struct}) also excludes other local minimum structure 
than the global equilibrium structure. 
Figure~\ref{fig:2PA-structure} illustrates this optimized molecular structure of 2-propanimine, 
in which we denote ``top 1'' and ``top 2'' for the two methyl tops hereinafter. 
``Top 1'' is the \ce{CH3} group pointing away from the \ce{N-H} bond, and ``top 2'' is the one facing to the \ce{N-H} bond.
Dipole moment calculation gives $\mu_a=1.37$~D and $\mu_b=-2.06$~D (MP2 value), which correspond to $a$- and $b$- type rigid rotor selection rules. 
We note that the dipole moment values reported by \cite{Sil2018ApJ} were erroneously taken from the initial structure 
of the molecule before geometry optimization, and therefore should not be used. 
We also show evidence that the normal $c$- type transitions, which correspond to a non-zero $\mu_c$ dipole moment component, 
do not exist in the experimental spectra. 
More technical analysis on the $c$- type selection rules are discussed in Appendix~\ref{app:ctype-sel-rule}.

\begin{figure}
  \centering
  \includegraphics[width=0.5\textwidth]{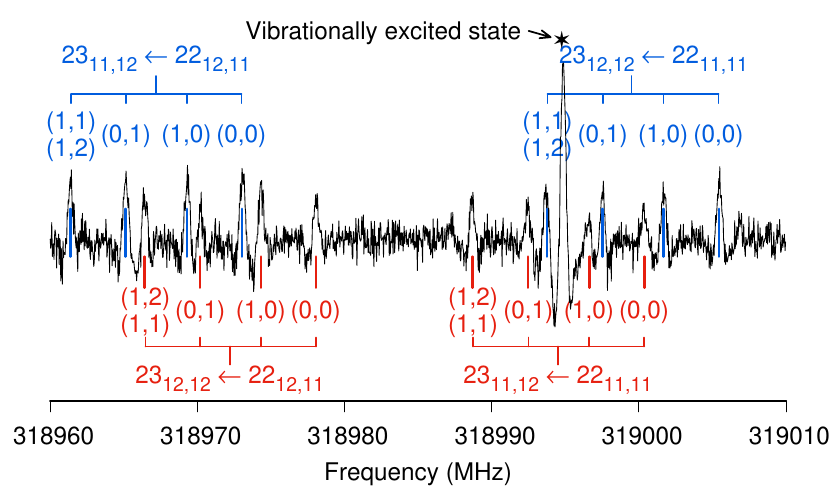}\\
  \includegraphics[width=0.5\textwidth]{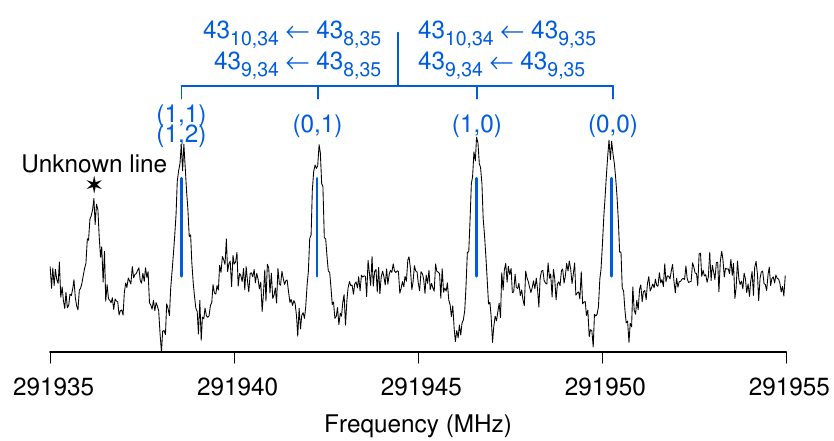}
  \caption{Spectrum of 2-propanimine with the characteristic quadruplets due to torsional splitting. 
  The vertical sticks represent the predicted frequencies of the transitions from the final ERHAM fit.
  $R$ branch lines are plotted on the top panel, and $Q$ branch lines are plotted on the bottom panel. 
  In the $R$ branch panel, sticks in blue represent $b$-type transitions, whereas sticks in pink represent $a$-type transitions. 
  In the $Q$ branch panel, $a$-type and $b$-type transitions become blended.
  \label{fig:sample-spectrum-normal}}
\end{figure}

In total, we identified 27,259 torsional-rotational transitions and 9,349 distinct line frequencies in our spectral analysis. 
The highest upper state $J$ and $K_a$ accessed are 52, 26 for $R$ branch lines, and 64, 34 for $Q$ branch lines.
The spectrum measured between 50 and 110 GHz was extremely useful because it accessed a large number of $Q$ branch transitions with high $J$ and $K_a$ values. 
All transition frequencies, uncertainties and relative weights used in the fit are listed in Table~\ref{tbl:linelist}.
Typical $a$-type and $b$-type $R$ and $Q$ lines are plotted against the predicted frequencies from the best fit in Figure~\ref{fig:sample-spectrum-normal}.

\begin{table*}
  \centering
  \caption{Measured and assigned transitions of 2-propanimine. \tablefootmark{(1)} \label{tbl:linelist}}
  \begin{tabular}{c c c c c c c c r r r r}
    \hline\hline
      $J''$ & $K_a''$ & $K_c''$ & $\sigma_1$ & $J'$ & $K_a'$ & $K_c'$ & $\sigma_2$ & $\nu_\text{obs}$ (MHz) & Uncertainty (MHz) & Weight \tablefootmark{(2)} & $\nu_\text{obs}-\nu_\text{calc}$ (MHz)\tablefootmark{(3)}\\
    \hline
      $3$ & $3$ & $1$ & $1$ & $2$ & $2$ & $0$ & $2$ & $53833.392$ & $0.050$ & $1.000$ & $ 0.0474$ \\
      $3$ & $3$ & $1$ & $1$ & $2$ & $2$ & $0$ & $1$ & $53834.919$ & $0.050$ & $1.000$ & $ 0.0397$ \\
      $3$ & $3$ & $1$ & $0$ & $2$ & $2$ & $0$ & $1$ & $53837.233$ & $0.025$ & $1.000$ & $ 0.0263$ \\
      $3$ & $3$ & $1$ & $1$ & $2$ & $2$ & $0$ & $0$ & $53840.874$ & $0.025$ & $1.000$ & $-0.0103$ \\
      $3$ & $3$ & $1$ & $0$ & $2$ & $2$ & $0$ & $0$ & $53843.939$ & $0.025$ & $1.000$ & $-0.0373$ \\
    \hline\hline
  \end{tabular}
  \vspace*{-1ex}
  \tablefoot{
    \tablefoottext{1}{The full table is available in machine-readable format in the online supplementary material. }
    \tablefoottext{2}{Weight used in fitting blended transitions. Weight is proportional to the line intensity and normalized to unity for all blended transitions associated with a single frequency.}
    \tablefoottext{3}{$\nu_\text{calc}$ obtained from the best ERHAM fit parameters (see Table~\ref{tbl:fit-result}).}
  }
\end{table*}

\begin{figure}
  \centering
  \includegraphics[width=0.45\textwidth]{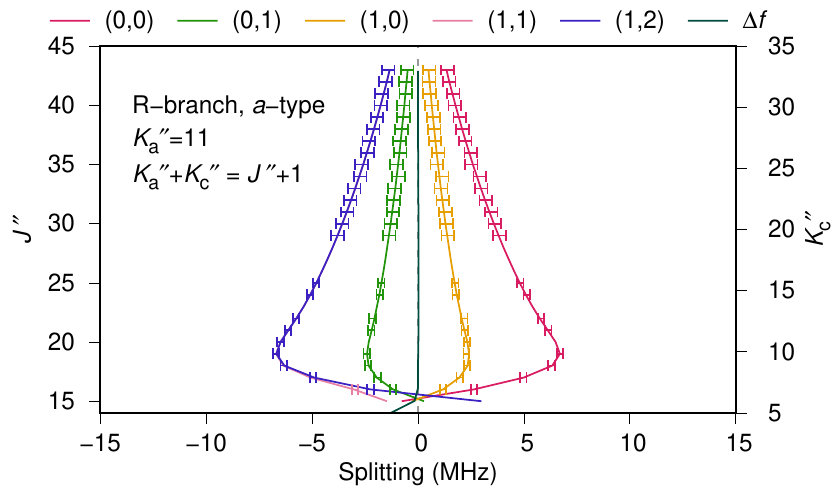}\\
  \includegraphics[width=0.45\textwidth]{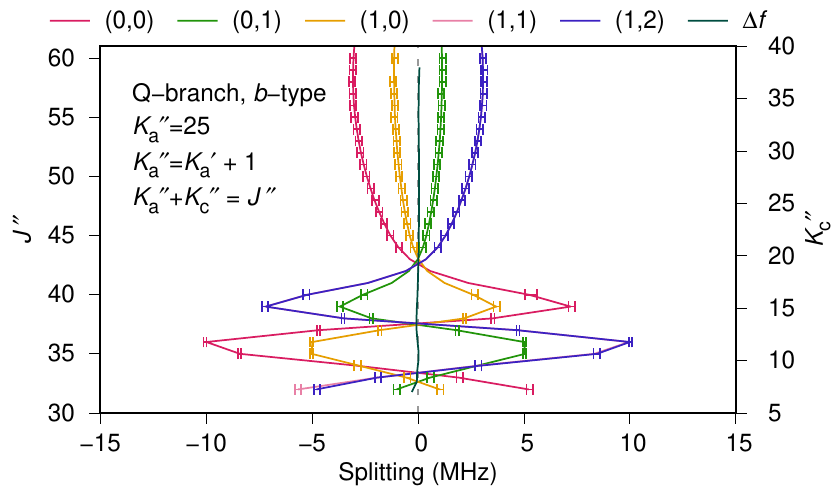}
  \caption{Torsional splitting patterns for (top) $a$-type, R-branch, $K_a''=11$, $K_a''+K_c''=J''+1$ series; and (bottom) $b$-type, Q-branch, $K_a''=25$, $K_a''=K_a'+1$, and $K_a''+K_c''=J''$. 
  The horizontal axis is the shift of each torsional splitting component from the center frequency $f_c$, which is the weighted average of the frequencies of all splitting components.
  The predicted frequency of each component is plotted as a solid curve, and the experimental values are marked as error bars with $\pm3$ times the assigned measurement uncertainty. 
  The dark green solid curve in the center is $\Delta f$, the difference between $f_c$ and its asymmetric rotor fit result. 
  \label{fig:split-pattern}}
\end{figure}

The fit results with XIAM and ERHAM  are both listed in Table~\ref{tbl:fit-result}.
The XIAM fit produces a microwave root-mean-square (r.m.s.) of 78.0~kHz and a dimensionless standard deviation of 1.38. 
The \mbox{ERHAM} fit produces a lower microwave r.m.s.\ of 53.5~kHz and a dimensionless standard deviation of 0.69. 
Detailed statistical analysis (see Appendix~\ref{app:fit}) shows that the ERHAM model reaches experimental accuracy,
whereas the XIAM model cannot reach experimental accuracy because of the limited number of high order correction terms available in the model. 
The goodness of the fit and the torsional splitting pattern can also be visualized in Figure~\ref{fig:split-pattern},
where we plot the torsional splitting among the five components as a function of the upper state $J''$ and $K_c''$. 
In this figure, the center frequency ($f_c$) of each $J''$ level is calculated from the weighted average of the observed line frequencies of the five torsional components.
Then the observed line frequencies are plotted as the deviation from this center frequency with $\pm 3\sigma$ error bars. 
The predicted frequencies from the ERHAM fit are plotted as continuous curves. 
These curves follow nicely with the observed frequencies within their uncertainty range.
For most $J''$ values, the $(1,1)$ and $(1,2)$ components are merged, leaving four components symmetric around the center frequency.
When $K_c''$ becomes sufficiently small, the splitting between $(1,1)$ and $(1,2)$ states becomes visible. 
In the figure, it is shown as the separation of the curve in pale pink (the $(1,1)$ component) from the curve in blue (the $(1, 2)$ component).
The torsional splitting pattern of the $R$ branch is relatively simple. 
The splitting is maximized at medium $J''$ value, and decreases with both the increase and decrease of $J''$. 
The torsional splitting pattern of the $Q$ branch is more complicated. 
As $J''$ decreases, the splitting first decreases to zero where the five components cross each other, and then the splitting increases again. 
This trend repeats three times to create three cross-overs of the five components. 
It is also interesting to examine the separation of the torsional splitting from the molecule's overall asymmetric rotation.  
If we assume that there were no torsional splitting from internal rotation, 
the imagined pure rotational transition of 2-propanimine would be the $f_c$.
We generated a line list from $f_c$ and fitted it to a simple Watson's A reduction Hamiltonian in $I^r$ representation using the CALPGM/SPFIT program \citep{Pickett1991JMS}.
Then, the frequency difference from the fit is denoted as $\Delta f$ and plotted as dark green solid curves in Figure~\ref{fig:split-pattern}.
The almost invisible $\Delta f$ shows that indeed an asymmetric rotor model is able to describe the averaged center frequencies $f_c$ of the torsional splitting components. 

\begin{table*}
  \centering
  \footnotesize
  \caption{Spectroscopic parameters of 2-propanimine obtained from XIAM and ERHAM fit and from theoretical calculation.\label{tbl:fit-result}}
  \begin{tabular}{p{2.0cm} l r r | p{2.1cm} l r r}
  \hline\hline
   Overall rotation parameters & Unit & XIAM & ERHAM & Tunneling parameters \tablefootmark{(1)} & Unit & \multicolumn{2}{c}{ERHAM}  \\
  \hline
    $      A      $ & MHz & $ 9709.08619(11) $ & $ 9709.082826(66) $ & $ \epsilon_{01}       $ & MHz & \multicolumn{2}{c}{$ -64.4590(86)  $} \\
    $      B      $ & MHz & $ 8479.04382(16) $ & $ 8478.033891(58) $ & $ \epsilon_{02}       $ & MHz & \multicolumn{2}{c}{$  -0.0226(36)  $} \\
    $      C      $ & MHz & $ 4788.51105(14) $ & $ 4789.524383(54) $ & $ \epsilon_{10}       $ & MHz & \multicolumn{2}{c}{$ -30.0609(79)  $} \\
    $ \Delta_{J}  $ & kHz & $   4.671205(73) $ & $    4.673412(56) $ & $ \epsilon_{12}       $ & MHz & \multicolumn{2}{c}{$   0.0204(33)  $} \\
    $ \Delta_{JK} $ & kHz & $   -5.47595(24) $ & $    -5.47338(14) $ &                         &     & $ qq'=01        $ & $ qq'=10        $ \\
    $ \Delta_{K}  $ & kHz & $   10.46953(39) $ & $    10.44279(23) $ & $ [g_a]_{qq'}         $ & MHz & $   0.04253(92) $ & $  -0.00335(92) $ \\
    $ \delta_{J}  $ & kHz & $   1.982110(24) $ & $    1.982158(15) $ & $ [D_{ab}]_{qq'}      $ & MHz & $   0.0444(85)  $ & $   0.0416(90)  $ \\
    $ \delta_{K}  $ & kHz & $   0.826549(90) $ & $    0.829265(63) $ & $ [A-(B+C)/2]_{qq'}   $ & kHz & $   3.671(67)   $ & $   2.042(66)   $ \\
    $ \Phi_{J}    $ &  Hz & $   0.005608(21) $ & $    0.006205(17) $ & $ [(B+C)/2]]_{qq'}    $ & kHz & $  -1.691(38)   $ & $  -1.146(41)   $ \\
    $ \Phi_{JK}   $ &  Hz & $   -0.00536(12) $ & $   -0.000357(69) $ & $ [(B-C)/4]_{qq'}     $ & kHz & $  -0.376(21)   $ & $  -0.353(23)   $ \\
    $ \Phi_{KJ}   $ &  Hz & $   -0.04460(40) $ & $    -0.06331(22) $ & $ [\Delta_J]_{qq'}    $ &  Hz & $   0.910(16)   $ & $   0.451(17)   $ \\
    $ \Phi_{K}    $ &  Hz & $    0.08289(52) $ & $     0.08478(33) $ & $ [\Delta_{JK}]_{qq'} $ &  Hz & $  -2.456(56)   $ & $  -1.085(60)   $ \\
    $ \phi_{J}    $ &  Hz & $  0.0027909(75) $ & $   0.0030980(46) $ & $ [\Delta_K]_{qq'}    $ &  Hz & $   1.667(56)   $ & $   0.620(59)   $ \\
    $ \phi_{JK}   $ &  Hz & $   0.011037(58) $ & $    0.014644(38) $ & $ [\delta_J]_{qq'}    $ &  Hz & $   0.4987(82)  $ & $   0.2390(86)  $ \\
    $ \phi_{K}    $ &  Hz & $   -0.00344(15) $ & $   -0.007352(88) $ & $ [\delta_K]_{qq'}    $ &  Hz & $  -1.042(16)   $ & $  -0.453(17)   $ \\
  \hline
  \end{tabular}
  \begin{tabular}{l l r r | r r}
  \hline
    \multirow{2}{2.2cm}{Internal rotation parameters} & Unit & \multicolumn{2}{c}{Top 1} & \multicolumn{2}{c}{Top 2} \\ 
    \cline{3-4} \cline{5-6}
    & & XIAM & ERHAM & XIAM & ERHAM \\
  \hline
    $ V_3 $                      & cm$^{-1}$ & $  531.956(64)  $ & $ 531.394 $ \tablefootmark{(2)} & $  465.013(26)  $ & $ 462.284 $ \tablefootmark{(2)} \\
    $ \rho $                     &           & $ 0.0597139(75) $ & $ 0.059652(20)      $   & $ 0.0591996(34) $ & $ 0.0589308(90) $ \\
    $ F $                        & cm$^{-1}$ & $ 5.60425       $ \tablefootmark{(3)} & $ 5.6016(20)        $   & $ 5.62699       $ \tablefootmark{(3)} & $ 5.64411(87)   $ \\
    $ \beta $ \tablefootmark{(4)}        & degree    & $  26.3833      $ \tablefootmark{(3)} & $ 27.032(23) $ & $ 151.6718 $ \tablefootmark{(3)} & $ 151.1511(91) $ \tablefootmark{(6)} \\
    $ \alpha $ \tablefootmark{(5)}       & degree    & $  0 $ \tablefootmark{(7)} & $ 0 $ \tablefootmark{(7)} & $ 0 $ \tablefootmark{(7)} & $ 0 $ \tablefootmark{(7)} \\
    $ \angle(i,a)$ \tablefootmark{(8)}   & degree    & $  29.5966(49)  $ & $ 30.299(24)        $   & $ 148.3136(24)  $ & $ 147.7540(97)  $ \\
    $ \angle(i,b)$ \tablefootmark{(8)}   & degree    & $  60.4034(49)  $ & $ 59.701(24)        $   & $  58.3136(24)  $ & $ 57.7540(97)   $ \\
    $ \angle(i,c)$ \tablefootmark{(8)}   & degree    & $  90.0         $\tablefootmark{(7)} & $ 90.0              $\tablefootmark{(7)}   & $  90.0         $\tablefootmark{(7)} & $ 90.0          $\tablefootmark{(7)} \\
    $ \Delta_{pi2J} $\tablefootmark{(9)} & MHz       & $  0.0372(38)   $ &                         & $  0.1351(18)   $ & \\
    $ \Delta_{pi2K} $\tablefootmark{(9)} & MHz       & $ -0.158(12)    $ &                         & $ -0.6502(58)   $ & \\
    $ \Delta_{pi2-} $\tablefootmark{(9)} & MHz       & $  0.0201(28)   $ &                         & $  0.0538(13)   $ & \\
  \hline\hline
    Fit Statistics & Unit & \multicolumn{2}{r}{XIAM} & \multicolumn{2}{r}{ERHAM} \\
  \hline
    $n$ \tablefootmark{(10)}                &     & &  9349 & &  9349 \\ 
    $N$ \tablefootmark{(11)}                &     & & 27259 & & 27259 \\ 
    $\sigma_\text{MW}$ \tablefootmark{(12)} & kHz & &  78.0 & &  53.5 \\ 
    $\sigma_w$ \tablefootmark{(13)}         &     & &  1.38 & &  0.69 \\ 
  \hline\hline 
  \end{tabular}
  \vspace*{-1ex}
  \tablefoot{
    \tablefoottext{1}{Parameters only available in ERHAM.}
    \tablefoottext{2}{Approximate values derived from program ``BARRIER'' using torsional energy differences. No statistical standard deviation is available because there is no degree of freedom in the data set. }
    \tablefoottext{3}{Values derived from XIAM and no  statistical standard deviation  is available. }
    \tablefoottext{4}{Polar angle between the $\rho$ vector axis and the $a$ principal axis.}
    \tablefoottext{5}{Azimuthal angle between the $\rho$ vector axis and the $b$ principal axis.}
    \tablefoottext{6}{In ERHAM, the angle in the input file is the supplementary angle of $\beta$.}
    \tablefoottext{7}{Fixed. }
    \tablefoottext{8}{Angle between the internal rotor symmetry axis and the $a$, $b$, or $c$ principal axis. }
    \tablefoottext{9}{Parameters only available in XIAM. }
    \tablefoottext{10}{Number of individual line frequencies.}
    \tablefoottext{11}{Number of assigned transitions. }
    \tablefoottext{12}{Microwave root-mean-square of the fit $[\sum((\nu_\text{obs}-\nu_\text{calc})/w)^2/\sum((1/w)^2)]^{1/2}$.}
    \tablefoottext{13}{Dimensionless standard deviation of the fit $[\sum((\nu_\text{obs}-\nu_\text{calc})/w)^2/N]^{1/2}$.}
  }
\end{table*}

Both XIAM and ERHAM fits determine well the parameters of the molecule's overall rotation, 
including the rotational constants and the full set of quartic and sextic centrifugal distortion constants.
The parameters from these two fits agree well with each other, except the $B$ and $C$ constants which differ by $\sim$1~MHz.
It is not surprising because the two programs treat the internal rotation differently, 
and therefore may introduce different ``effective'' rotational constants introduced by the internal rotors into the $B$ and $C$ constants. 
For the internal rotation part, we note that XIAM and ERHAM are two different types of models, 
and therefore they do not share the exact same set of parameters. 
XIAM fits directly the $V_3$ barrier, the $\rho$ parameter, 
the angle between the internal rotor axis and the $\rho$ axis of the molecule ($\beta$),
and the three internal rotational correction parameters $\Delta_{pi2J}$, $\Delta_{pi2K}$, and $\Delta_{pi2-}$.
The $V_3$ barriers to internal rotation of the two rotors are determined to be 531.956(64)~cm$^{-1}$ and 465.013(26)~cm$^{-1}$, respectively.
ERHAM also fits the $\rho$ and the angle parameters, but it does not directly fit the barriers to internal rotation. 
Instead, it fits the tunneling parameters $\epsilon_{01}$, $\epsilon_{10}$, etc., and the correction terms for each individual rotor. 
The $V_3$ barriers are derived from the above parameters \citep{Groner1986JMS}, 
and the results are 531.394~cm$^{-1}$ and 462.284~cm$^{-1}$, respectively.
These values are lower than the XIAM fitted values by less than 3~cm$^{-1}$.
The $\rho$ and $\angle(i,a)$ determined by the two fits also agree well with each other. 

\begin{figure}
  \centering
  \includegraphics[width=0.45\textwidth]{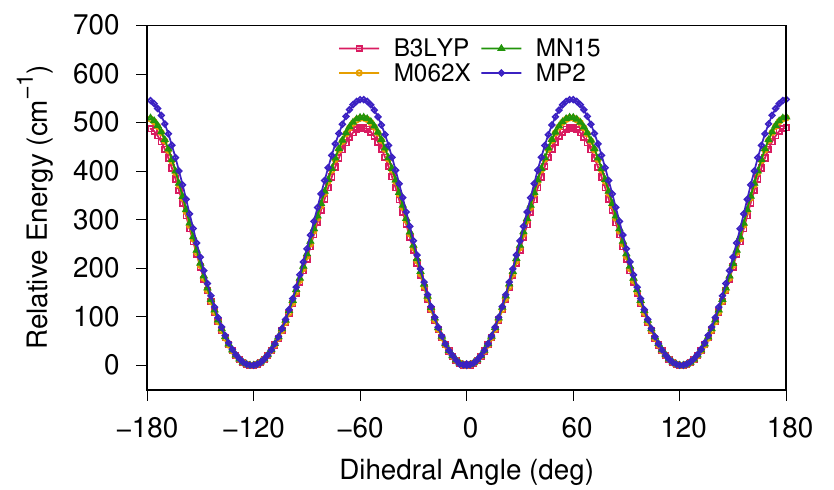} \\
  \includegraphics[width=0.45\textwidth]{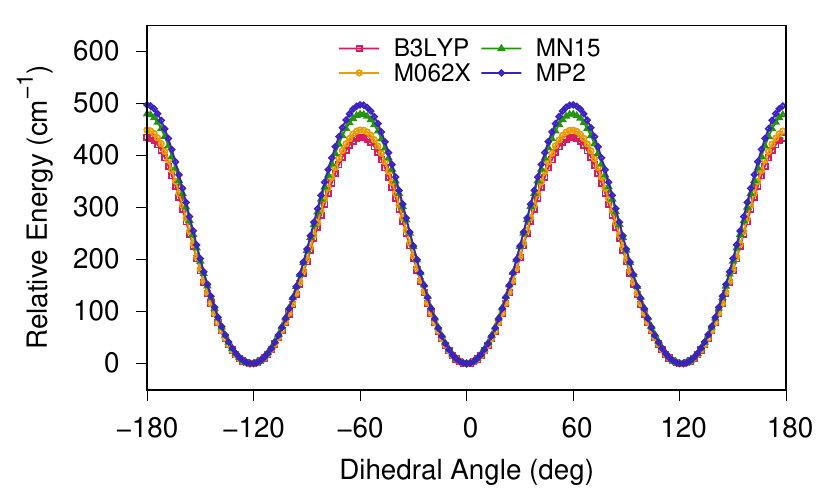}
  \caption{The 1-dimension PES of ``top 1'' (top panel) and ``top 2'' (bottom panel) \ce{CH3} internal rotation in 2-propanimine. \label{fig:2PA-pes-DFT}}
\end{figure}

We compare several molecular structure parameters and the barriers to internal rotation of the two methyl tops from the spectral fit with theoretical calculation values in Table~\ref{tbl:fit-theory}. 
The calculation systematically over-predicts the $A, B, C$ rotational constants with $<$1.5~\% error. 
MP2 predicts the most accurate $A$ constant, whereas B3LYP predicts the most accurate $B$ and $C$ constants. 
For the $V_3$ barriers, Figure~\ref{fig:2PA-pes-DFT} shows the 1-dimensional PES scans using different level of theories. 
Overall, the calculated $V_3$ barriers derived from the PES scans agree with the experimentally fitted values within 10~\% difference. 
MP2 and MN15 return values closer to experimentally fitted values than B3LYP or M062X. 
MP2 overestimates the barriers, whereas DFT methods underestimate the barriers.
MN15 is an outlier in predicting the energy difference of the two barriers as the difference is significantly underestimated compared to the other methods.
All calculations predict similar moment of inertia of the \ce{CH3} tops $I_\alpha$, which is smaller than the experimentally fitted value. 
The difference is expected because the experiment fits the effective structure, whereas calculation predicts equilibrium structures that lack the contribution from vibrational deformation. 
The smaller calculated $I_\alpha$ values also lead to higher calculated $\rho$. 
The angle between the internal rotation axis and $a$ principal axis is systematically underestimated by a few degrees. 
Based on these comparisons, we may consider that the MP2/aug-cc-pVTZ method predicts the parameters with the smallest averaged error to experimental fitted values.
It is interesting that ``top 2'', the \ce{CH3} top closer to the \ce{N-H} bond, has the lower $V_3$ barrier. 

\begin{table*}
  \centering
  \footnotesize
  \caption{Comparison of molecular structure parameters between fit and calculation. \label{tbl:fit-theory}}
  \begin{tabular}{l l r r r r r r}
    \hline\hline
      Parameter & Unit & ERHAM fit & XIAM fit & B3LYP & M062X & MN15 & MP2 \\
    \hline
      $ A $                   & MHz             & $9709.082826(66) $ & $9709.08619(11) $ & $ 9757.1877 $ & $ 9780.7341 $ & $ 9771.8994 $ & $ 9721.3053 $ \\
      $ B $                   & MHz             & $8478.033891(58) $ & $8479.04382(16) $ & $ 8508.6190 $ & $ 8573.1854 $ & $ 8584.7372 $ & $ 8559.6800 $ \\
      $ C $                   & MHz             & $4789.524383(54) $ & $4788.51105(14) $ & $ 4813.2847 $ & $ 4839.8416 $ & $ 4841.0024 $ & $ 4821.2849 $ \\
      $ V_3  $        (top 1) & cm$^{-1}$       & $ 531.394        $ & $ 531.956(64)   $ & $ 489.49    $ & $ 494.03    $ & $ 511.30    $ & $ 547.64    $ \\
      $ V_3  $        (top 2) & cm$^{-1}$       & $ 462.284        $ & $ 465.013(26)   $ & $ 435.52    $ & $ 432.16    $ & $ 479.90    $ & $ 497.47    $ \\
      $ \rho $        (top 1) &                 & $ 0.059652(20)   $ & $ 0.0597139(75) $ & $ 0.0602179 $ & $ 0.0604033 $ & $ 0.0603622 $ & $ 0.0600987 $ \\
      $ \rho $        (top 2) &                 & $ 0.0589308(90)  $ & $ 0.0591996(34) $ & $ 0.0599404 $ & $ 0.0601071 $ & $ 0.0600893 $ & $ 0.0598149 $ \\
      $ I_{\alpha} $  (top 1) & amu$\cdot\text{\r{A}}^2$ & $ 3.2034(12)     $ & $ 3.20231(40)$  & $ 3.123812  $ & $ 3.125991  $ & $ 3.126738  $ & $ 3.129077  $ \\
      $ I_{\alpha} $  (top 2) & amu$\cdot\text{\r{A}}^2$ & $ 3.17675(52)    $ & $ 3.18760(19)$  & $ 3.109383  $ & $ 3.110658  $ & $ 3.112605  $ & $ 3.114300  $ \\
      $ \angle(i,a) $ (top 1) & degree          & $ 30.299(24)     $ & $ 29.5966(49)   $ & $ 25.756    $ & $ 25.482    $ & $ 24.612    $ & $ 26.627    $ \\
      $ \angle(i,a) $ (top 2) & degree          & $ 147.7540(97)   $ & $ 148.3136(24)  $ & $ 152.131   $ & $ 151.704   $ & $ 150.794   $ & $ 152.821   $ \\
      $ \mu_a $ \tablefootmark{(1)} & Debye        &                 &                   & $  1.3216   $ & $  1.3142   $ & $  1.2797   $ & $  1.3711   $ \\    
      $ \mu_b $ \tablefootmark{(1)} & Debye        &                 &                   & $ -2.1122   $ & $ -2.1066   $ & $ -2.1242   $ & $ -2.0643   $ \\    
      $ \nu_{27} $ \tablefootmark{(2)} & cm$^{-1}$ &                 &                   & $ 122.447   $ & $ 114.190   $ & $ 126.645   $ & $ 119.892   $ \\
      $ \nu_{26} $ \tablefootmark{(2)} & cm$^{-1}$ &                 &                   & $ 178.662   $ & $ 183.843   $ & $ 185.232   $ & $ 183.404   $ \\
    \hline\hline 
  \end{tabular}
  \vspace*{-1ex}
  \tablefoot{
    \tablefoottext{1}{The vibrational averaged dipole moment at 298 K.}
    \tablefoottext{2}{Harmonic vibrational energy.}
  }
\end{table*}

Using the ERHAM fit, we generated a prediction of the transition frequencies of 2-propanimine up to $J''=80$ and 1~THz.
We used the dipole moments from the MP2 calculation to calculate the line strength.
The predicted transition frequencies, along with their uncertainties derived from the uncertainties of fitted parameters, 
are listed in Table~\ref{tbl:2PA-pred} and also available in machine-readable format at CDS (reference J/MNRAS/Vol/Page) 
and also at the Lille Spectroscopic Database\footnote{https://lsd.univ-lille.fr/} \citep{Motiyenko2022ISMS-LSD}.
Nevertheless, the predicted frequencies of transitions not accessed by this experiment should still be viewed with caution because the high order terms in the ERHAM fit may be highly correlated and 
subject to large fluctuation if more spectral data were accessible. 
The torsional-rotational partition function $Q(T)$ was calculated by direct summation of Boltzmann factors up to $J''=120$ (maximum $J$ allowed in ERHAM) and listed in Table~\ref{tbl:partition}. 
We note that the partition function and the upper state degeneracy listed in Table~\ref{tbl:2PA-pred} are correlated and affected by the choice of spin statistics.
If we consider the full spin weight of the 6 H-atoms in the two methyl tops, the total spin weight is $2^{6}=64$.
The upper state degeneracies of the $(\sigma_1, \sigma_2) = (0,0),(0,1)(1,0)(1,1)(1,2)$ torsional states will then be 16, 16, 16, 8, 8 times the rigid rotor state degeneracy, respectively. 
On the other hand, the CALPGM/SPCAT format of the prediction uses 3 columns to label the upper state degeneracy, 
and therefore has a plafond of 999. 
For high $J$ states, the consideration of full spin statistics leads to degeneracies exceeding 999.
For this practical reason, we used a reduced spin weight of 2:2:2:1:1 for the five torsional states. 
The total spin weight reduces from 64 to 8, and therefore the partition function also reduces to 1/8 of the full spin weight case.  
The logarithm line strength remains unchanged. 
To correctly calculate line intensities using our prediction, it is necessary to choose the correct pair of spin weight and partition function. 
Furthermore, vibrational partition function can be estimated using the 
vibrational energies listed in Table~\ref{tbl:AEvib}, and combined with the torsional-rotational partition function 
to provide more accurate line intensity estimation under elevated temperature. 

\begin{table*}
  \centering
  \caption{Prediction of 2-propanimine transitions up to $J''=80$ and 1~THz using the ERHAM fit and reduced spin weight. 
  Prediction uses dipole moment values from the MP2/aug-cc-pVTZ calculation, and the table follows the CALPGM/SPCAT format.\tablefootmark{(1)} \label{tbl:2PA-pred}}
  \begin{tabular}{c c c c r r c c r r r r r r r r}
    \hline\hline
    FREQ & UNC & LOGINT\tablefootmark{(2)} & DR & ELO & GUP\tablefootmark{(3)} & TAG & QNFMT & $J''$ & $K_a''$ & $K_c''$ & $\sigma_1$ & $J'$ & $K_a'$ & $K_c'$ & $\sigma_2$ \\
    \hline
    $17162.6416$ & $0.0009$ & $-6.5439$ & $3$ & $ 20.1322$ & $34$ & $57803$ & $1404$ & $8$ & $6$ & $2$ & $0$ & $8$ & $5$ & $3$ & $1$ \\
    $17165.2573$ & $0.0010$ & $-6.5437$ & $3$ & $ 20.1294$ & $34$ & $57803$ & $1404$ & $8$ & $6$ & $2$ & $1$ & $8$ & $5$ & $3$ & $0$ \\
    $17168.1874$ & $0.0011$ & $-6.5436$ & $3$ & $ 20.1269$ & $34$ & $57803$ & $1404$ & $8$ & $6$ & $2$ & $0$ & $8$ & $5$ & $3$ & $0$ \\ 
    \hline\hline
  \end{tabular}
  \vspace*{-1ex}
  \tablefoot{
    \tablefoottext{1}{The full table is available in machine-readable format in the online supplementary material, as well as the CDS database (reference J/MNRAS/Vol/Page) and Lille Spectroscopic database.}
    \tablefoottext{2}{Intensity calculated using reduced partition function $Q=353614.251$ at 300~K.  }
    \tablefoottext{3}{To get full spin weight of 6-H atoms, multiply this degeneracy by 8. The partition function should also be multiplied by 8 to get correct line intensity calculation. }
  }
\end{table*}

\begin{table}
  \centering
  \caption{Rotational partition function of 2-propanimine from direct summation of Boltzmann factors up to $J=120$. \label{tbl:partition}}
  \begin{tabular}{l r r}
    \hline\hline
      Temperature (K) & $Q_\text{full}$\tablefootmark{(1)}  & $Q_\text{red}$ \tablefootmark{(2)}\\
    \hline
      300    & 2828914.006 & 353614.251 \\
      225    & 1836813.509 & 229601.689 \\
      150    &  999569.239 & 124946.155 \\
       75    &  353416.309 &  44177.039 \\
       37.5  &  125041.237 &  15630.155 \\
       18.75 &   44286.338 &   5535.792 \\
        9.38 &   15727.703 &   1965.963 \\
        5.00 &    6160.952 &    770.119 \\
    \hline\hline
  \end{tabular}
  \vspace*{-1ex}
  \tablefoot{
    \tablefoottext{1}{Partition function with full spin weight of 6-H atoms, total weight $2^{6}=64$.}
    \tablefoottext{2}{Partition function with reduced spin weight, total weight $2^{3}=8$. }
  }
\end{table}

Harmonic vibrational frequency calculations predict that two torsional excited states lie around 100--200~cm$^{-1}$ above the ground state. 
By examining the derivative of normal coordinates, we can associate these two excited states to the torsional motion of the two methyl tops. 
The exact values of state energies are subject to some uncertainty due to the level of theory used in the calculation.
If we take the MP2/aug-cc-pVTZ value based on its best agreement of overall rotation and internal rotation parameters to experimentally fitted values, 
the energy of the two torsional excited states are 120~cm$^{-1}$ and 183~cm$^{-1}$, respectively, or 134~cm$^{-1}$ and 171~cm$^{-1}$ with anharmonic correction. 
They correspond to 40--60~\% of the ground state population under 300~K. 
Under typical interstellar temperature ($\sim$150--200~K for hot cores), 
the intensity of the lines from these excited states are expected to be less than 30~\% of the ground state lines.
The search for these excited state lines may help the identification of the molecule 
when ground state lines are heavily blended with lines from other molecules, 
if the ground state lines themselves are sufficiently bright. 
In our laboratory spectra, we observed numerous lines from these two states with intensities consistent with our expectations.
The analysis of these excited states, however, is complicated due to their Coriolis coupling.
To correctly model the lines from these excited states, 
it is necessary to perform a global analysis of them together with the ground state lines.
The global analysis, however, cannot be treated straightforwardly with the tools we used to analyze the ground state. 
In this regard, we focus only on the analysis of and astronomical search for the ground state lines in the current manuscript. 
The global spectroscopic modeling of these interacting torsional excited states 
and the search for these lines in the ISM will be the subject of future study. 

\section{Search for 2-propanimine in the ISM}

\label{s:astro}

\subsection{Search toward Sgr~B2(N1)}
\label{ss:obs_remoca}

We used the imaging spectral line survey Reexploring Molecular Complexity with
ALMA (ReMoCA) that targeted the high-mass star forming protocluster Sgr~B2(N) 
with ALMA. Details about the observations and data reduction can 
be found in \citet{Belloche19}. We summarize here the main features of the 
survey. The phase center is located at the equatorial position 
($\alpha, \delta$)$_{\rm J2000}$= 
($17^{\rm h}47^{\rm m}19{\fs}87, -28^\circ22'16{\farcs}0$). This position is 
half-way between the two hot molecular cores Sgr~B2(N1) and Sgr~B2(N2). We
covered the frequency range from 84.1~GHz to 114.4~GHz at a spectral 
resolution of 488~kHz (1.7 to 1.3~km~s$^{-1}$) using five different frequency 
tunings. The survey achieved a sensitivity per spectral channel that varies 
between 0.35~mJy~beam$^{-1}$ and 1.1~mJy~beam$^{-1}$ (rms) depending on the 
setup, with a median value of 0.8~mJy~beam$^{-1}$. The observations have an 
angular resolution (HPBW) ranging from $\sim$0.3$\arcsec$ to 
$\sim$0.8$\arcsec$ with a median value of 0.6$\arcsec$ that corresponds to 
$\sim$4900~au at the distance of Sgr~B2 \citep[8.2~kpc,][]{Reid19}. We used 
here an improved version of the data reduction, as described in 
\citet{Melosso20}.

We followed the same strategy as \citet{Belloche19} and analyzed the spectrum 
obtained toward the position Sgr~B2(N1S) at ($\alpha, \delta$)$_{\rm J2000}$= 
($17^{\rm h}47^{\rm m}19{\fs}870$, $-28^\circ22\arcmin19{\farcs}48$). This 
position is offset by about 1$\arcsec$ to the south of the main hot core 
Sgr~B2(N1) and has a lower continuum opacity compared to 
the peak of the hot core. The observed spectrum was compared to synthetic 
spectra computed under the assumption of local thermodynamic 
equilibrium (LTE) with the astronomical software Weeds \citep[][]{Maret11}. 
This assumption is justified by the high densities of the regions where 
hot-core emission is detected in Sgr~B2(N) \citep[$>1 \times 10^{7}$~cm$^{-3}$, 
see][]{Bonfand19}. We derived by hand a best-fit synthetic spectrum 
for each molecule 
separately, and then added together the contributions of all identified 
molecules. Each species was modeled with a set of five parameters: size of the 
emitting region ($\theta_{\rm s}$), column density ($N$), temperature 
($T_{\rm rot}$), linewidth ($\Delta V$), and velocity offset ($V_{\rm off}$) 
with respect to the assumed systemic velocity of the source, 
$V_{\rm sys}=62$~km~s$^{-1}$. 

The molecules included in the complete model comprise in particular 
those listed in \citet{Belloche13}. The linewidths and velocity offsets are 
directly evaluated on the individual lines. As explained in 
\citet{Belloche19}, the molecular emission toward Sgr~B2(N1S) is resolved and 
we fixed its size to 2$\arcsec$ for the LTE modeling. This size is much larger 
than the beam and, therefore, the determination of the column densities does 
not depend on its exact value. The optimization of the rotation temperatures 
is guided by population diagrams \citep[see][]{Belloche19}. In the end, for a 
given molecule, the only really free parameter is the column density, which we 
adjust until the peak temperatures of the synthetic lines match the detected 
ones.

%\subsection{Nondetection of 2-propanimine}
%\label{ss:nondetection_remoca}

\begin{figure}
\centering
\includegraphics[width=0.45\textwidth]{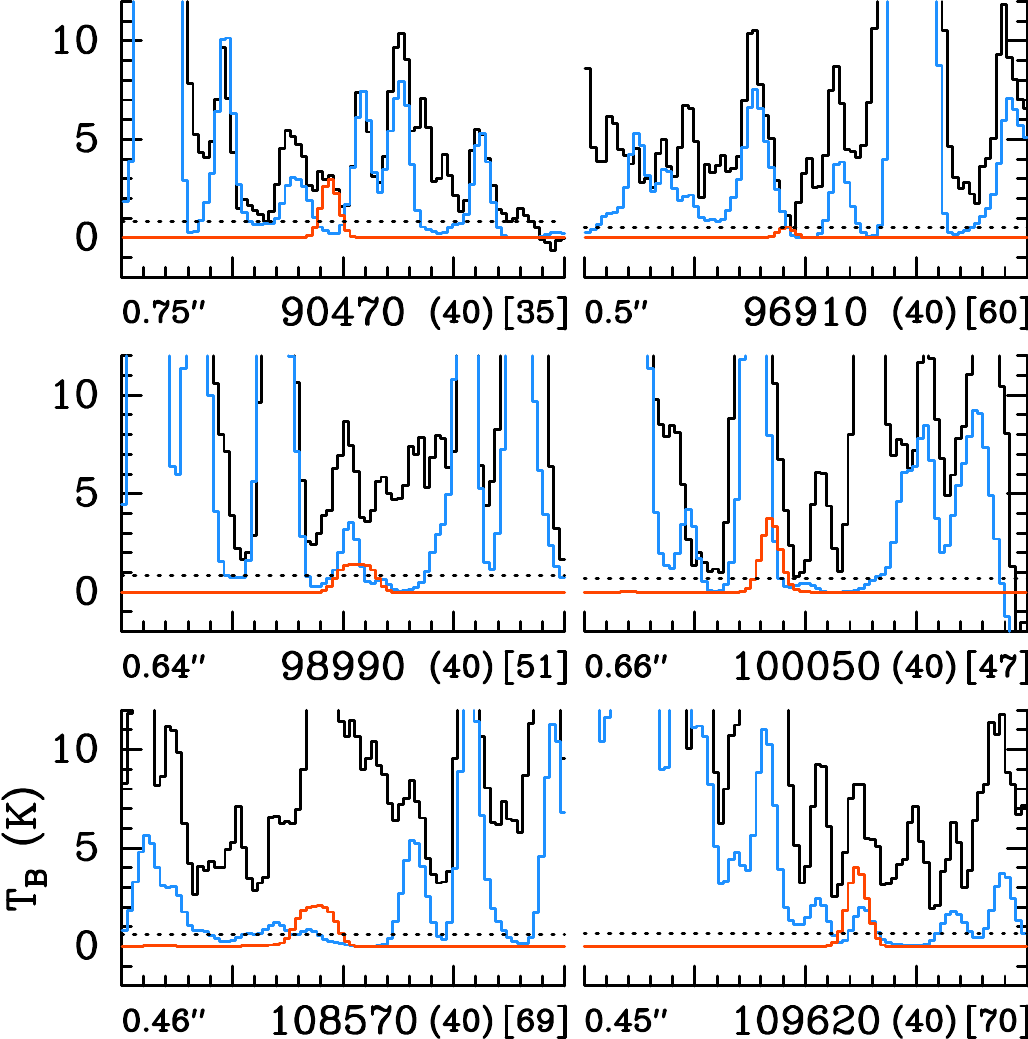}
\caption{Selection of transitions of 2-propanimine CH$_3$C(NH)CH$_3$ covered 
by the ReMoCA survey. The LTE synthetic spectrum used to derive the upper limit
on the column density of CH$_3$C(NH)CH$_3$ is displayed in red and overlaid on 
the observed spectrum of Sgr~B2(N1S) shown in black. The blue synthetic 
spectrum contains the contributions of all molecules identified in our survey 
so far, but does not include the contribution of the species shown in red. The 
values written below each panel correspond from left to right to the half-power 
beam width, the central frequency in MHz, the width in MHz of each panel in 
parentheses, and the continuum level in K of the baseline-subtracted spectra 
in brackets. The y-axis is labeled in brightness temperature units (K). The 
dotted line indicates the $3\sigma$ noise level.}
\label{f:remoca_propanimine-2}
\end{figure}

In order to search for 2-propanimine, CH$_3$C(NH)CH$_3$, toward Sgr~B2(N1S),
we relied on the LTE parameters that we previously derived for methanimine, 
CH$_2$NH, toward this position with the ReMoCA survey \citep[][]{Margules22}.
We employed the spectroscopic predictions derived for 2-propanimine in 
Sect.~\ref{sect:result} to compute LTE synthetic spectra and search for emission of 
this molecule. Figure~\ref{f:remoca_propanimine-2} illustrates the results of
this search. The spectrally unresolved multiplet of transitions of 
2-propanimine at $\sim$90.469~GHz matches a line detected in the ReMoCA 
spectrum (top left panel of Fig.~\ref{f:remoca_propanimine-2}). No other 
molecule contributes to the emission at this frequency in our current complete 
model of Sgr~B2(N1S). However, all other transitions of 2-propanimine that are 
expected to be stronger than $3\sigma$ are heavily blended with emission from 
other molecules and cannot be identified. The spectrally unresolved multiplet 
of transitions of 2-propanimine at $\sim$109.625~GHz
(bottom right panel of Fig.~\ref{f:remoca_propanimine-2}) contributes to about
half of the flux density of the line detected at this frequency, with another
$\sim$25\% contributed by an isotopolog of ethyl cyanide. This is not 
sufficient to secure the identification of the observed line. Overall, although
the data may hint at the presence of 2-propanimine in Sgr~B2(N1S), the 
evidence is not robust enough to claim a detection of this molecule, not even a
tentative detection. Therefore, we consider the synthetic spectrum shown in red
in Fig.~\ref{f:remoca_propanimine-2} as an upper limit to the emission of 
2-propanimine in this source and we report the corresponding upper limit on 
its column density in Table~\ref{t:coldens}, after accounting for the 
vibrational partition function. To our knowledge, 2-propanimine does not have 
conformers, which means that no conformational correction to the column 
density is needed.

We recall in Table~\ref{t:coldens} the column density of methanimine and the
upper limit to the column densities of the \textit{Z} and \textit{E} conformers 
of 1-propanimine that we derived earlier with the ReMoCA survey toward 
Sgr~B2(N1S) \citep[][]{Margules22}. Table~\ref{t:coldens} indicates that 
propanimine is at least 18 times less abundant than methanimine in 
Sgr~B2(N1S). The upper limit is a factor three more stringent than the one
obtained previously for 1-propanimine.

\begin{table*}
  \begin{center}
  \caption{
  Parameters of our best-fit LTE model of methanimine toward Sgr~B2(N1S), and upper limits for 1-propanimine and 2-propanimine.
 }
  \label{t:coldens}
  \vspace*{-1.2ex}
  \begin{tabular}{lcrcccccccr}
  \hline\hline
  \multicolumn{1}{c}{Molecule} & \multicolumn{1}{c}{Status\tablefootmark{a}} & \multicolumn{1}{c}{$N_{\rm det}$\tablefootmark{b}} & \multicolumn{1}{c}{$\theta_{\rm s}$\tablefootmark{c}} & \multicolumn{1}{c}{$T_{\mathrm{rot}}$\tablefootmark{d}} & \multicolumn{1}{c}{$N$\tablefootmark{e}} & \multicolumn{1}{c}{$F_{\rm vib}$\tablefootmark{f}} & \multicolumn{1}{c}{$F_{\rm conf}$\tablefootmark{g}} & \multicolumn{1}{c}{$\Delta V$\tablefootmark{h}} & \multicolumn{1}{c}{$V_{\mathrm{off}}$\tablefootmark{i}} & \multicolumn{1}{c}{$\frac{N_{\rm ref}}{N}$\tablefootmark{j}} \\ 
   & & & \multicolumn{1}{c}{\small ($''$)} & \multicolumn{1}{c}{\small (K)} & \multicolumn{1}{c}{\small (cm$^{-2}$)} & & & \multicolumn{1}{c}{\small (km~s$^{-1}$)} & \multicolumn{1}{c}{\small (km~s$^{-1}$)} & \\ 
  \hline
  \textit{Methanimine} & & & & & & & & & \\[0.ex] 
  \hspace*{2ex} CH$_2$NH\tablefootmark{(k)}$^{(\star)}$ & d & 4 &  2.0 &  230 &  9.0 (17) & 1.00 & -- & 5.0 & 0.0 &       1 \\ 
 \hline 
  \textit{1-Propanimine} & & & & & & & & & \\[0.ex] 
  \hspace*{2ex} \textit{E}-C$_2$H$_5$CHNH\tablefootmark{(k)} & n & 0 &  2.0 &  230 & $<$  1.5 (17) & 3.54 & 1.19 & 5.0 & 0.0 & $>$     6.1 \\ 
  \hspace*{2ex} \textit{Z}-C$_2$H$_5$CHNH\tablefootmark{(k)} & n & 0 &  2.0 &  230 & $<$  2.0 (17) & 3.58 & 6.22 & 5.0 & 0.0 & $>$     4.5 \\ 
 \hline 
  \textit{2-Propanimine} & & & & & & & & & \\[0.ex] 
  \hspace*{2ex} CH$_3$C(NH)CH$_3$ & n & 0 &  2.0 &  230 & $<$  5.0 (16) & 3.33 & 1.00 & 5.0 & 0.0 & $>$      18 \\ 
 \hline 
  \end{tabular}
  \end{center}
  \vspace*{-2.5ex}
  \tablefoot{
  \tablefoottext{a}{d: detection, n: nondetection.}
  \tablefoottext{b}{Number of detected lines \citep[conservative estimate, see Sect.~3 of][]{Belloche16}. One line of a given species may mean a group of transitions of that species that are blended together.}
  \tablefoottext{c}{Source diameter (FWHM).}
  \tablefoottext{d}{Rotational temperature.}
  \tablefoottext{e}{Total column density of the molecule. $x$ ($y$) means $x \times 10^y$. For 1-propanimine, the two conformers were modeled as independent species and a conformer correction ($F_{\rm conf}$) was applied a posteriori, such that each column density corresponds to the total column density of 1-propanimine.}
  \tablefoottext{f}{Correction factor that was applied to the column density to account for the contribution of vibrationally excited states, in the cases where this contribution was not included in the partition function of the spectroscopic predictions.}
  \tablefoottext{g}{Correction factor that was applied to the column density to account for the contribution of other conformers in the cases where this contribution was not included in the partition function of the spectroscopic predictions.}
  \tablefoottext{h}{Linewidth (FWHM).}
  \tablefoottext{i}{Velocity offset with respect to the assumed systemic velocity of Sgr~B2(N1S), $V_{\mathrm{sys}} = 62$ km~s$^{-1}$.}
  \tablefoottext{j}{Column density ratio, with $N_{\rm ref}$ the column density of the previous reference species flagged with a star ($^\star$).}
  \tablefoottext{k}{The parameters were derived from the ReMoCA survey by \citet{Margules22}.}
  }
\end{table*}
 
\subsection{Search toward IRAS 16293-2422}

\label{jj:astro}

We also searched for 2-propanimine toward the ``B component'' of the low-mass protostar IRAS~16293-2422 
from data obtained in connection with the Protostellar Interferometric Line Survey (PILS) program \citep{Jorgensen2016AA}. 
We refer to \citet{Jorgensen2016AA} for details about the survey but repeat the key information needed for this analysis.
PILS is an unbiased line survey of the Class~0 protostellar system, IRAS~16293-2422, often considered an astrochemical template source. 
PILS covers the frequency range from 329.1~GHz to 362.9~GHz in ALMA's Band~7 at 0.2~km~s$^{-1}$ spectral resolution. 
We focus on a position offset by one beam (0.5$''$ or 70~au) from the B component where the lines are narrow ($\approx$~1~km~s$^{-1}$) 
and absorption due to optical thickness (of continuum and line) is less than toward the location of the source itself. 
Toward this position methanimine \ce{CH2NH} was detected as part of PILS by \citet{Ligterink2018AA} 
with a column density of 6--10$\times 10^{14}$~cm$^{-2}$ for excitation temperatures between 70 and 120~K.

\begin{figure}
\centering
\includegraphics[width=0.48\textwidth]{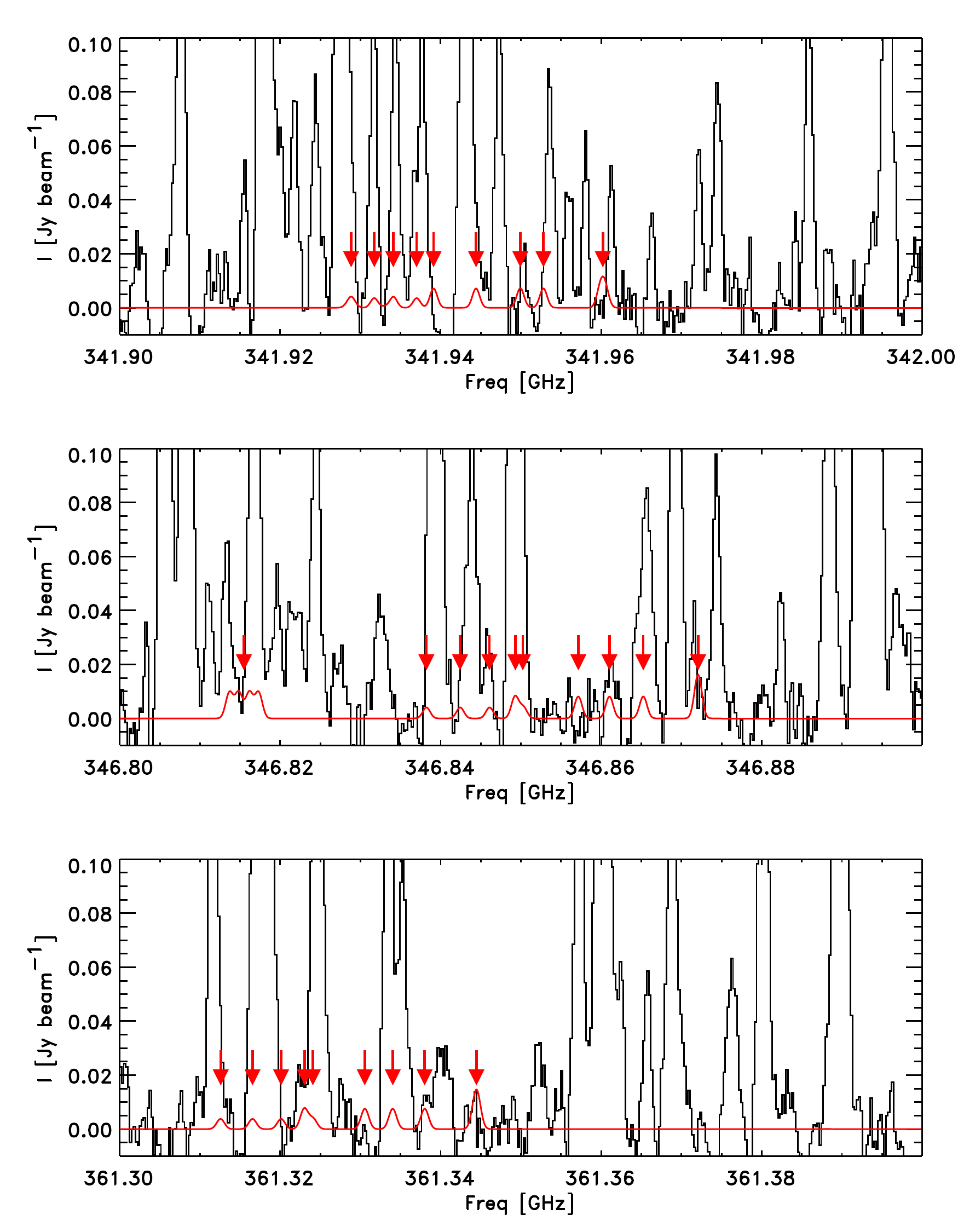}
\caption{Search for 2-propanimine toward IRAS~16293B: the shown frequency windows contain some of the brightest transitions 
predicted for an excitation temperature of 100~K. No lines can be claimed to be detected at this level and 
the upper limit (red line) corresponds to a column density of $5\times 10^{14}$~cm$^{-2}$.\label{jj:fig_iras}}
\end{figure}

%\subsection{Non detection of 2-propanimine in IRAS 16293B}
To perform the search, we calculated synthetic spectra for 2-propanimine assuming optically thin emission and LTE
using custom routines, which are used also for other papers from the PILS program \citep[see for example][]{Jorgensen2016AA, Ligterink2018AA, Calcutt2018A90, Manigand2020A48, Coutens2022A6}
but equivalent to the methodology of using Weeds in the search for 2-propanimine toward Sgr B2(N1S).
We adopted a temperature of 100~K and systemic velocities consistent with other species detected in PILS. 
The spectral regions with the brightest predicted lines in the PILS range are shown in Fig.~\ref{jj:fig_iras}. 
The upper limit to column density can be derived by comparing to the parts of the spectra 
where no lines are seen down to the RMS noise level of the data (4--5 mJy~beam$^{-1}$~km~s$^{-1}$)
but where 2-propanimine is predicted to show emission, e.g., around 341.96, 346.86 and 361.34~GHz. 
The upper limit found in this manner is $5\times10^{14}$~cm$^{-2}$, i.e., a ratio < 0.5--1 with respect to \ce{CH2NH},
thus significantly less constraining than the < 1/18 ratio toward Sgr B2(N1S).

\section{Conclusions}

We have presented the measurement and analysis of the millimeter-wave spectrum of 2-propanimine between 50 and 500~GHz for the first time. 
We successfully assigned and fitted 27,259 transitions and 9,349 individual frequencies of 2-propanimine using XIAM and ERHAM, 
and fully determined its spectroscopic parameters and the barriers to internal rotation of the two nonequivalent \ce{CH3} tops in 2-propanimine. 
A fit residual of 53.5~kHz and dimensionless standard error of 0.69 were obtained by ERHAM using a total of 44 parameters, 
including the full set of rotational constants, quartic and sextic centrifugal distortion constants, and 12 internal rotation and tunneling parameters. 
The barriers to internal rotation are fitted experimentally to 533.764(63)~cm$^{-1}$ and 466.657(26)~cm$^{-1}$ by XIAM. 
The barriers and the molecular structure parameters from the fit agree well with theoretical calculation values. 
A prediction to $J''=80$ and 1~THz is generated using the best-fit parameters from ERHAM. 

We report the nondetection of 2-propanimine toward the offset position Sgr~B2(N1S) of the hot molecular core Sgr~B(N1) and toward the protostar IRAS~16293B. 
We find that 2-propanimine is at least 18 times less abundant than methanimine in Sgr~B2(N1), 
and at most 50--83~\% of methanimine in IRAS~16293B.

\section*{Acknowledgements}

The authors thank Zbigniew Kisiel for his helpful discussion and modification of the ERHAM program, 
and Isabelle Kleiner and Lam Nguyen for their reminder of a technical point in the XIAM fit.
This work was supported by the CNES and the Action sur Projets de l'INSU, PCMI, 
and by the European Union's Horizon 2020 research and innovation programme 
under the Marie Sk{\l}odowska-Curie grant agreement (H2020-MSCA-IF-2019, Project no. 894508).
This paper makes use of the following ALMA data: ADS/JAO.ALMA\#2016.1.00074.S, ADS/JAO.ALMA\#2013.1.00278.S.
ALMA is a partnership of ESO (representing its member states), NSF (USA), and 
NINS (Japan), together with NRC (Canada), NSC and ASIAA (Taiwan), and KASI 
(Republic of Korea), in cooperation with the Republic of Chile. The Joint ALMA 
Observatory is operated by ESO, AUI/NRAO, and NAOJ. The interferometric data 
are available in the ALMA archive at https://almascience.eso.org/aq/.
Part of this work has been carried out within the Collaborative
Research Centre 956, sub-project B3, funded by the Deutsche
Forschungsgemeinschaft (DFG) -- project ID 184018867.

\section*{Data Availability}

The data underlying this article are available in the online supplementary material of this article, 
as well as in the Zenodo repository (Appendix~\ref{app:c}, DOI: 10.5281/zenodo.7541890).
The spectral line catalog is also available at CDS via anonymous ftp to cdsarc.u-strasbg.fr (130.79.128.5) 
or via https://cdsarc.unistra.fr/viz-bin/cat/J/MNRAS,
and the Lille Spectroscopic Database (https://lsd.univ-lille.fr/).

\bibliographystyle{mnras}
\bibliography{2022_2PA}

\newpage

\appendix

\section{Experimental details}\label{app:exp}

\subsection{Synthesis of 2-propanimine}
2-Amino-2-methylpropanenitrile, prepared as previously reported \citep{Jenny1986HelvChim}, 
was vaporized in a vacuum line equipped with a reactor half-filled 
with \ce{KOH} in powder form and heated to 90~\textdegree{}C.
The vacuum line was also equipped with two U-tubes with stopcocks \citep[for similar experiments see][]{Guillemin2019CC}.
The first stopcock was immersed in a cold bath at $-60$~\textdegree{}C to trap compounds with high boiling point,
and the second one in a bath cooled to $-100$~\textdegree{}C to selectively trap the 2-propanimine. 
At the end of the reaction, the stopcocks of the second U-tube were closed,
and the 2-propanimine sample was preserved in dry ice before connecting to the cell of the spectrometer.

\subsection{Millimeter-wave spectroscopy of 2-propanimine}

The spectrum was measured using the fast absorption spectrometer at Lille. 
The details of the spectrometer have been described elsewhere \citep{Zakharenko2015JMS,Motiyenko2019AA}, 
and here we only state the specific conditions concerning this measurement. 
After connecting the sample U-tube to our 2-meter-long glass absorption cell, the dry ice bath was removed, 
and the U-tube was immediately submerged into an ethanol bath at $-60$~\textdegree{}C throughout the measurement.
The stopcock was slightly opened to introduce a small continuous flow of the sample vapor into the cell. 
The cell pressure was maintained between 8--20~$\upmu$Bar, optimized in each millimeter-submillimeter wave band for the best signal-to-noise ratio (SNR). 

The spectrum was measured between 50--110~GHz, 150--330~GHz, and 360--500~GHz
using commercial frequency amplifier-multiplier chains (50--75~GHz: Millitech AMC-15-R0000, above 75~GHz: Viginia Diodes, Inc.) driven by a microwave synthesizer (Agilent, E8527). 
Signals were detected by solid state detectors (Viginia Diodes, Inc.).
We applied sine-wave frequency modulation and second harmonics detection using a lock-in amplifier (METEK 7270 DSP).
For the spectrum above 150~GHz, the modulation deviation after frequency multiplication 
was set to match the linewidth of the spectral lines, 
at 240~kHz in 150--220 GHz, 270~kHz in 225--330~GHz, and 540~kHz in 360--500~GHz. 
The time constant was set to 0.2~ms, and 2--4 acquisitions were sufficient.
For the spectrum below 110~GHz, the spectral intensity is weak due to the unfavorable Boltzmann distribution at room temperature.
To improve the SNR, we slightly exaggerated the modulation deviation over the spectral linewidth.
The modulation deviation was set to 200~kHz in 50--75~GHz, and 300~kHz in 75--110~GHz.
A longer time constant, 1~ms in 50--75~GHz, and 0.5~ms in 75--110~GHz, was also chosen together with 8 acquisitions to obtain the spectra with decent SNR. 

\subsection{Quantum chemical calculation}

Quantum chemical calculations were performed on the Sakura high-performance computer cluster at PhLAM, using the Gaussian16 software \citep{Gaussian16}.
Molecular structure of 2-propanimine, fixed to $C_s$ symmetry, was optimized, and harmonics vibrational frequencies were calculated based on the optimized geometry. 
Several density functional theory (DFT) methods, B3LYP \citep{B3LYP}, M062X \citep{M062X}, MN15 \citep{MN15}, were used with the 6-311++G(3df,3pd) basis set, which include polarization and diffuse functions.  
Second order M\o{}ller–Plesset perturbation (MP2) calculation was used with aug-cc-pVTZ basis set. 
The results of these calculations were used to estimate the barrier to internal rotation of the two \ce{CH3} groups in 2-propanimine, 
and to guide the assignment of the experimental spectra.
To estimate the barrier to internal rotation of the two \ce{CH3} groups, 
1-dimensional potential energy surface (PES) scan was conducted by twisting the dihedral angle of one of the \ce{C-H} bond in the \ce{CH3} group with respect to the central \ce{C-C(N)-C} plane. 
In addition, 1-dimensional PES scan was performed on the angle of the \ce{N-H} bond with respect to the central \ce{C-C(N)-C} plane, in order to verify the symmetry group of the molecule.
During the PES scans, the molecular symmetry was relaxed to allow re-optimization of the molecular structure at each scanned point. 

\subsection{Spectral analysis}

The experimental spectra were processed using a custom spectral assignment software. 
In the first step, the sinusoidal baseline produced by standing waves in the absorption cell was removed by Fourier-transforming the spectrum, erasing their representative frequency components, 
and then inverse Fourier-transforming back to the original frequency space. 
After baseline removal, peaks were identified and fitted with a second-derivative Voigt line profile to obtain their peak frequencies.
We associated various uncertainties, ranging from 25~kHz to 200~kHz, to each fitted peak according to the frequency resolution at each frequency band, 
and the SNR of the peaks represented by the peak frequency uncertainty from the least-square fit. 
Typical uncertainty of individual lines is 25~kHz in 50--110~GHz, 50~kHz in 150--330~GHz, and 100~kHz above 360~GHz, 
and we doubled the uncertainty for blended lines and lines with low SNR. 
Weights were also assigned to blended transitions according to their relative line intensities. 

The spectral fit was performed using both the XIAM \citep{Hartwig1996ZNatA, Herbers2020JMS} and ERHAM \citep{Groner1997JCP,Groner2012JMS} programs 
with the standard Watson's A reduction Hamiltonian under the $I^r$ representation.
As there is no available microwave study of 2-propanimine to our knowledge, we started our assignment using XIAM to predict the spectrum of 2-propanimine 
using the rotational constants, quartic centrifugal distortion constants, and internal rotation parameters obtained from the theoretical calculation. 
Once the strongest $R$-branch lines were identified and assigned, the assignment and fit was performed iteratively by including more series of weaker lines into the fit. 
Throughout this iterative process, we used Loomis-Wood diagrams \citep{LoomisWood1928PhysRev} to assist the identification of line series.
Finally, ERHAM was used to provide a refined fit with higher accuracy using the same input line list as for the XIAM fit. 

We noted that the number of transitions in our data set exceeded the input limit of the publicly available XIAM and ERHAM programs, 
hosted at Zbigniew Kisiel's website PROSPE\footnote{http://www.ifpan.edu.pl/$\sim$kisiel/prospe.htm} (Programs for ROtational SPEctroscopy) \citep{Kisiel2001Prospe}.
Therefore, the source codes of both programs were slightly modified to allow larger number of input lines. 
For XIAM, we obtained the source code of the official ``XIAM\_mod'' program\footnote{http://www.ifpan.edu.pl/$\sim$kisiel/introt/xiam\_mod/XIAM\_mod.zip}, 
and modified the ``\verb|DIMLIN|'' parameter, which controls the limit of input lines, to 99999. 
We also changed the output format of the fitted parameters to display more significant figures for the sextic centrifugal distortion constants. 
The modified source code was compiled using the original MAKEFILE in the source code package on a Windows 10 computer using the GNU Fortran 11.1.0 compiler.
For ERHAM, the limit of input lines was increased to 32768. 
The modification was kindly provided by Dr.~Kisiel, and the modified ERHAM code, named \verb|erham_r3a.exe|, can be also found in Kisiel's website%
\footnote{http://www.ifpan.edu.pl/$\sim$kisiel/introt/erham/erhamz\_R3a.exe}.

\begin{figure}
  \centering
  \includegraphics[width=0.48\textwidth]{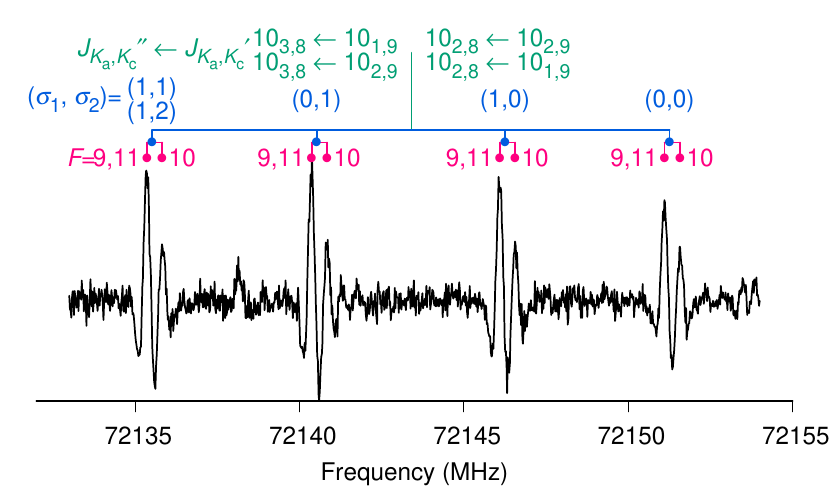}
  \caption{Example of the treatment of the partially resolved hyperfine components. 
  The observed hyperfine component frequencies are marked with pink dots and sticks. 
  The weighted averaged frequencies of these components, marked in blue dots and sticks, were used in the spectral fit.\label{fig:hfs-average}}
\end{figure}

Quadrupole hyperfine splitting due to the \ce{^{14}N} nuclei was partially resolved in the spectra at low frequencies. 
Only the three strongest components, namely, the $\Delta F=\Delta J$ components, were observable under our sensitivity level.
The $F=J\pm1$ components are blended and the $F=J$ component is separated. 
Therefore, the spectral line of each torsional-rotational transition further splits into 2 lines with intensity ratio of roughly 2:1. 
The full treatment of the coupling between the nuclear spin with the internal rotor is extremely complicated \citep{Van2020JMStr}. 
Without the access to microwave-cavity spectrum at higher resolution, our millimeter-wave data alone were not sufficient to fully determine this interaction. 
In addition, the number of partially resolved hyperfine lines consists of only a small portion, about 1~\%, of the complete line list.
Therefore, instead of including the hyperfine effect into the spectroscopy model, 
we chose to treat these partially blended hyperfine lines separately before the final spectral fit. 
For each set of hyperfine components, we calculated the weighted averaged line frequency using the theoretical hyperfine intensity ratio as the weighting factor.
For $R$ branch lines, the intensity is proportional to 
\begin{equation*}
\frac{1}{F''}(J''+F''+2)(J''+F''+1)(J''+F''-1)(J''+F''-2),
\end{equation*}
and for $Q$ branch lines, the intensity is proportional to 
\begin{equation*}
\frac{2F''+1}{F''(F''+1)}\big(J''(J''+1) + F''(F''+1) - 2\big)^2.
\end{equation*}
$J''$ and $F''$ denote the $J$ and $F$ quantum numbers of the upper state associated with the transition \citep{TownsSchawlow}. 
These averaged line frequencies were then merged with other torsional-rotation line frequencies for the spectral fit. 
One example of such weighted average treatment is shown in Figure~\ref{fig:hfs-average}. 
Pink dots mark the experimentally measured frequencies of the hyperfine components, 
and blue dots mark their weighted average frequencies which were then used in the spectral fit. 

The barriers to internal rotation in the ERHAM model were derived from the torsional energy differences based on the method 
described in \cite{Groner1986JMS}.
The ``BARRIER'' program\footnote{http://www.ifpan.edu.pl/\~{}kisiel/introt/barrier/BARRIER.exe} available on the PROSPE website 
is designed to calculate such barrier to internal rotation for single methyl rotor. 
Generally, it does not treat the case of two methyl rotors.
In our particular fit, however, the statistically meaningful tunneling parameters are only those associated with the individual tops, i.e., $\epsilon_{01}, \epsilon_{02}$, and $\epsilon_{10}$, $\epsilon_{12}$. 
The cross terms such as $\epsilon_{1,-1}$ and $\epsilon_{1,1}$ are not significant. 
In this case, the two methyl tops are almost independent (for the ground state), 
and we can approximately use the ``BARRIER'' program to calculate their $V_3$ barriers individually. 
Since only the ground state energy difference is available, we calculate the $V_3$ barrier height from only one parameter; 
there is no estimated uncertainty.
From these terms ERHAM calculates the torsional energy differences, 
which can then be used to derive the barrier to internal rotation 
The $V_3$ barriers are determined to be 531.394~cm$^{-1}$ and 462.284~cm$^{-1}$, respectively.

\section{Supplementary results}

\subsection{Calculation results}\label{app:calc-xyz}

The optimized $x,y,z$ coordinates of 2-propanimine using different level of theory mentioned in Tables~\ref{tbl:A1}--\ref{tbl:A4}.
The harmonic and anharmonic vibrational energies of 2-propanimine calculated by MP2/aug-cc-pVTZ is listed in Table~\ref{tbl:AEvib}, 
which can be used to estimate the vibrational partition function. 

\begin{table}
  \centering
  \caption{Optimized geometry of 2-propanimine using B3LYP/6-311++G(3df,3pd). Unit is in \AA.\label{tbl:A1}}
  \begin{tabular}{c r r r}
  \hline
  Atom & $x$ & $y$ & $z$ \\
  \hline
  N & $ 0.045853$ & $ 1.430604$ & $ 0.000000$ \\
  C & $ 0.000000$ & $ 0.160884$ & $ 0.000000$ \\
  H & $ 1.009860$ & $ 1.765233$ & $ 0.000000$ \\
  C & $-1.340658$ & $-0.522810$ & $ 0.000000$ \\
  H & $-2.139230$ & $ 0.213603$ & $ 0.000000$ \\
  H & $-1.442464$ & $-1.166615$ & $ 0.876838$ \\
  H & $-1.442464$ & $-1.166615$ & $-0.876838$ \\
  C & $ 1.203943$ & $-0.750007$ & $ 0.000000$ \\
  H & $ 2.134212$ & $-0.184332$ & $ 0.000000$ \\
  H & $ 1.189704$ & $-1.401953$ & $-0.876286$ \\
  H & $ 1.189704$ & $-1.401953$ & $ 0.876286$ \\
  \hline
\end{tabular}
\end{table}

\begin{table}
  \centering
  \caption{Optimized geometry of 2-propanimine using M062X/6-311++G(3df,3pd). Unit is in \AA.\label{tbl:A2}}
  \begin{tabular}{c r r r}
  \hline
  Atom & $x$ & $y$ & $z$ \\
  \hline
  N & $ 0.047787$ & $ 1.428742$ & $ 0.000000$ \\
  C & $ 0.000000$ & $ 0.163482$ & $ 0.000000$ \\
  H & $ 1.014348$ & $ 1.754234$ & $ 0.000000$ \\
  C & $-1.337039$ & $-0.522386$ & $ 0.000000$ \\
  H & $-2.134648$ & $ 0.213414$ & $ 0.000000$ \\
  H & $-1.430683$ & $-1.164272$ & $ 0.877171$ \\
  H & $-1.430683$ & $-1.164272$ & $-0.877171$ \\
  C & $ 1.198484$ & $-0.750195$ & $ 0.000000$ \\
  H & $ 2.129394$ & $-0.188104$ & $ 0.000000$ \\
  H & $ 1.174546$ & $-1.398798$ & $-0.876535$ \\
  H & $ 1.174546$ & $-1.398798$ & $ 0.876535$ \\
  \hline
\end{tabular}
\end{table}

\begin{table}
  \centering
  \caption{Optimized geometry of 2-propanimine using MN15/6-311++G(3df,3pd). Unit is in \AA.\label{tbl:A3}}
  \begin{tabular}{c r r r}
  \hline
  Atom & $x$ & $y$ & $z$ \\
  \hline
  N & $ 0.062586$ & $ 1.429206$ & $ 0.000000$ \\
  C & $ 0.000000$ & $ 0.161241$ & $ 0.000000$ \\
  H & $ 1.032955$ & $ 1.748995$ & $ 0.000000$ \\
  C & $-1.341104$ & $-0.511567$ & $ 0.000000$ \\
  H & $-2.132354$ & $ 0.231737$ & $ 0.000000$ \\
  H & $-1.441746$ & $-1.153897$ & $ 0.876584$ \\
  H & $-1.441746$ & $-1.153897$ & $-0.876584$ \\
  C & $ 1.190014$ & $-0.759265$ & $ 0.000000$ \\
  H & $ 2.124642$ & $-0.201788$ & $ 0.000000$ \\
  H & $ 1.163343$ & $-1.409020$ & $-0.876051$ \\
  H & $ 1.163343$ & $-1.409020$ & $ 0.876051$ \\
  \hline
\end{tabular}
\end{table}

\begin{table}
  \centering
  \caption{Optimized geometry of 2-propanimine using MP2/aug-cc-pVTZ. Unit is in \AA.\label{tbl:A4}}
  \begin{tabular}{c r r r}
  \hline
  Atom & $x$ & $y$ & $z$ \\
  \hline
  N & $ 0.037113$ & $ 1.437086$ & $ 0.000000$ \\
  C & $ 0.000000$ & $ 0.156237$ & $ 0.000000$ \\
  H & $ 1.008541$ & $ 1.754321$ & $ 0.000000$ \\
  C & $-1.334190$ & $-0.529263$ & $ 0.000000$ \\
  H & $-2.132032$ & $ 0.207629$ & $ 0.000000$ \\
  H & $-1.429035$ & $-1.170936$ & $ 0.877544$ \\
  H & $-1.429035$ & $-1.170936$ & $-0.877544$ \\
  C & $ 1.203722$ & $-0.746205$ & $ 0.000000$ \\
  H & $ 2.129794$ & $-0.174356$ & $ 0.000000$ \\
  H & $ 1.187395$ & $-1.394967$ & $-0.877149$ \\
  H & $ 1.187395$ & $-1.394967$ & $ 0.877149$ \\
  \hline
\end{tabular}
\end{table}

\begin{table}
  \centering
  \caption{Harmonic and anharmonic vibrational energy of the fundamental bands of 2-propanimine, calculated by MP2/aug-cc-pVTZ.\label{tbl:AEvib}}
  \begin{tabular}{c r r}
  \hline
  Mode & $E_\text{harm}$ (cm$^{-1}$) & $E_\text{anharm}$ (cm$^{-1}$) \\
  \hline
  $\nu_{ 1}$ &  3453.022 &  3287.214 \\
  $\nu_{ 2}$ &  3193.330 &  3055.883 \\
  $\nu_{ 3}$ &  3168.510 &  3029.200 \\
  $\nu_{ 4}$ &  3067.467 &  2963.991 \\
  $\nu_{ 5}$ &  3061.911 &  2961.916 \\
  $\nu_{ 6}$ &  1694.215 &  1653.353 \\
  $\nu_{ 7}$ &  1496.457 &  1453.421 \\
  $\nu_{ 8}$ &  1488.849 &  1444.107 \\
  $\nu_{ 9}$ &  1427.083 &  1385.101 \\
  $\nu_{10}$ &  1409.339 &  1371.320 \\
  $\nu_{11}$ &  1355.848 &  1322.955 \\
  $\nu_{12}$ &  1135.544 &  1104.881 \\
  $\nu_{13}$ &  1089.884 &  1065.154 \\
  $\nu_{14}$ &   942.033 &   924.229 \\
  $\nu_{15}$ &   824.252 &   806.235 \\
  $\nu_{16}$ &   505.878 &   503.780 \\
  $\nu_{17}$ &   381.089 &   386.913 \\
  $\nu_{18}$ &  3143.859 &  3005.790 \\
  $\nu_{19}$ &  3139.602 &  3003.191 \\
  $\nu_{20}$ &  1508.360 &  1451.230 \\
  $\nu_{21}$ &  1485.739 &  1443.779 \\
  $\nu_{22}$ &  1112.528 &  1084.408 \\
  $\nu_{23}$ &  1049.536 &  1019.205 \\
  $\nu_{24}$ &   830.510 &   814.042 \\
  $\nu_{25}$ &   463.680 &   461.770 \\
  $\nu_{26}$ &   183.404 &   171.152 \\
  $\nu_{27}$ &   119.892 &   133.962 \\
  \hline
  \end{tabular}
\end{table}

\subsection{Molecular Structure}\label{app:mol-struct}

\begin{figure}
\centering
\includegraphics[width=0.48\textwidth]{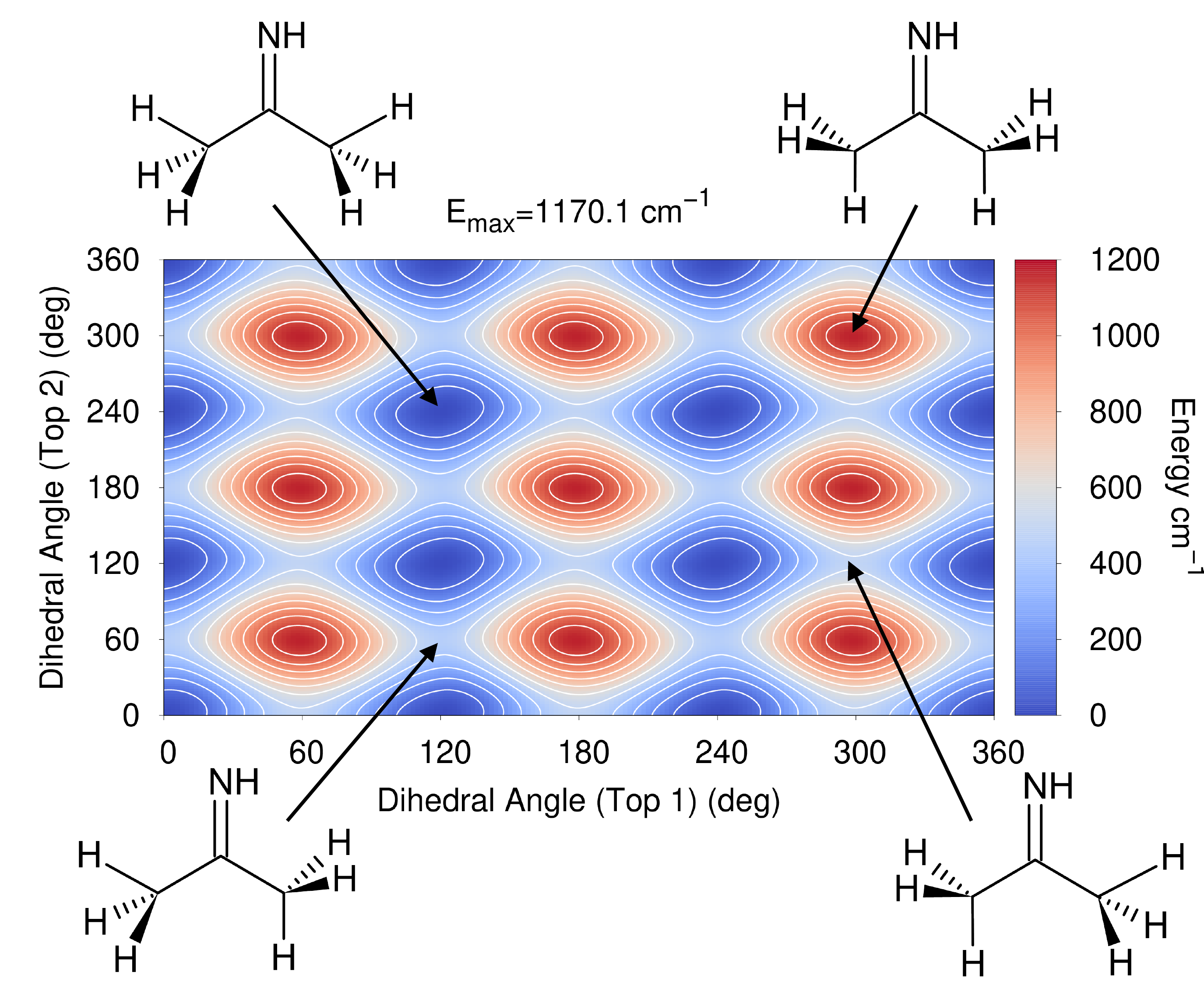}
\caption{2D PES of \ce{CH3} internal rotation in 2-propanimine calculated with B3LYP/6-311+g(2d,2p).
Energy step between contours is 100~cm$^{-1}$. The structure at the global minimum, maximum and 
two saddle points are presented and pointed by arrows. \label{fig:2D-PES}}
\end{figure}

Supplementary 2D PES scan regarding the internal rotation of \ce{CH3} tops
was performed to ensure that there is no other conformer, 
which are local minimum in the PES, except for the optimized conformation shown in Figure~\ref{fig:2PA-pes-DFT}.
The PES scan was performed from 0\textdegree{} to 120\textdegree{} with a step of 2\textdegree{}, and extrapolated to 360\textdegree{} based on the symmetry of the PES.
The basis set is reduced to 6-311+g(2d,2p) to speed up the calculation. 
The result is plotted in Figure~\ref{fig:2D-PES}. 
The $x$ and $y$ axes correspond to the dihedral angle of the rotation of each \ce{CH3} top. 
From the figure, We can find 9 equivalent global minima that correspond to the optimized 
conformation, and there is no other local minimum. 

\subsection{Selection rule and dipole moments}\label{app:ctype-sel-rule}

\begin{figure}
  \includegraphics[width=0.48\textwidth]{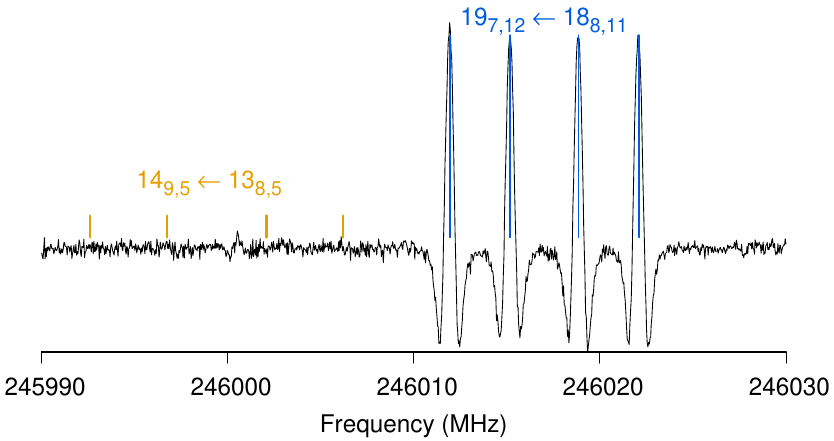}
  \caption{Non-detection of regular $c$-type transitions of 2-propanimine. The blue sticks represent the torsional splitting components of the $b$-type 
  transition $19_{7,12}\leftarrow 18_{8,11}$. The brown sticks represent the expected location of the $c$-type transition $14_{9,5}\leftarrow 13_{8,5}$, 
  if 2-propanimine would have $\mu_c=1$~D. The missing spectral lines proves that 2-propanimine does not possess 
  $\mu_c$, in agreement with its $C_s$ molecular symmetry. \label{fig:spectra-dipole-moment}}
\end{figure}

We have stated in Section~\ref{sect:result} that the dipole moment values of 2-propanimine 
we obtained from theoretical calculation ($\mu_a=1.37$~D and $\mu_b=-2.06$~D) 
differ from the values reported by \citeauthor{Sil2018ApJ} 
($\mu_a=-0.8107$~D, $\mu_b=1.8357$~D, and $\mu_c=1.4047$~D).
Several evidences point out that our values are more reliable and the values from \citeauthor{Sil2018ApJ} are erroneous. 
First, considering the $C_s$ symmetry of 2-propanimine, 
one of the three dipole moment components must be zero,
otherwise the molecule will have no symmetry plane. 
Secondly, selection rules of rotational transitions state that 
a non-zero $\mu_c$ dipole moment component corresponds to $c$- type selection rule, 
which is $\Delta K_c$ = even and $\Delta K_a$ = odd.
In our experimental spectra, we do no see these transitions.  
For example, in Figure~\ref{fig:spectra-dipole-moment}, 
we unambiguously show the missing of $c$-type transition lines $14_{9,5}\leftarrow 13_{8,5}$,
using the best-fit parameters from ERHAM and assumed $\mu_c=$1~D dipole moment.
Based on the noise level of the spectrum, the upper limit of the $\mu_c=$ is only 0.2~D. 
Thirdly, line intensity is proportional to the square of the dipole moment component. 
Using our dipole moment values, the intensity ratio between the $a$- and $b$- type transitions is 1:2.26.
If we use the dipole moment values from \citeauthor{Sil2018ApJ}, 
the intensity ratio between the $a$- and $b$- type transitions will be 1:5.13.
In Figure~\ref{fig:sample-spectrum-normal}, 
we have shown an example that the relative intensity of $a$- and $b$- type lines 
is closer to 1:2, which agrees better with our dipole moment values than the values from \citeauthor{Sil2018ApJ}.
With these reasons, we claim that our spectral data do not support the existence of $\mu_c$ dipole moment component in 2-propanimine, 
and the values from \citeauthor{Sil2018ApJ} should not be used. 

Despite the above arguments, we do have observed some spectral lines with $c$-type selection rule. 
This is not relevant to the dipole moment components, however, but an anomalous phenomenon caused by internal rotation.
The phenomenon has been observed and discussed in similar two methyl top systems, such as dimethyl ether \citep{Endres2009AA} and ethyl methyl ether \citep{Fuchs2003ApJS}. 
For a rigid rotor with a symmetry plane spanned by the $a$ and $b$ principal axes, only $a$- and $b$-type rotational transitions are allowed, e.g. transitions with $\Delta K_c$ = odd in the $J$, $K_a$, $K_c$ notation for the asymmetric rotor levels. 
For molecules with one methyl rotor, however, this selection rule from dipole moment restriction applies only to the $A$-state ($\sigma=0$) 
but not to the $E$-state ($\sigma=1$). 
For molecules with two methyl rotors, it applies only to the $(\sigma_1, \sigma_2)=(0,0)$ state,
but not to any state where $\sigma_1$ or $\sigma_2$ or both are non-zero. 
In these cases, the $J$, $K_a$, $K_c$ labels are at best approximate and at worst useless. 
Unfortunately, no easy-to-apply contradiction-free systems to assign asymmetric rotor labels exist in these cases. 
However, when such labels are used anyway 
(e.g., when the levels ordered by increasing energy are assigned as if they were those of a rigid asymmetric rotor), 
it occurs frequently that, (based on the labels,) some transitions with $\Delta K_c$ = even (instead of $\Delta K_c$ = odd) 
``appear'' regardless of $\mu_c=0$. 
This effect is due to the labeling problem, and ``$d$-type'' transitions (with $\Delta K_a = \Delta K_c$ = even) may also appear. 
Due to heavy mixing of asymmetric rotor wavefunctions, 
it is also common that both $b$-type and ``$c$-type'' (or $a$-type and ``$d$-type'') transitions may be observed instead of one or the other.

\begin{figure}
  \includegraphics[width=0.49\textwidth]{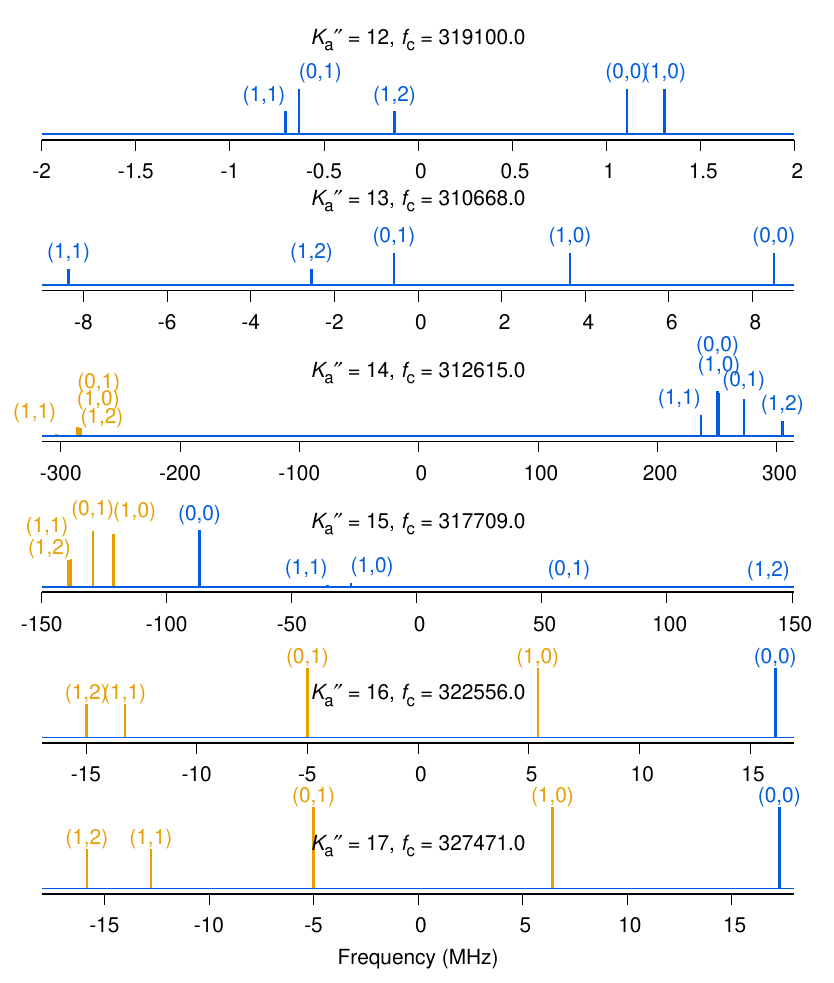}
  \caption{Demonstration of the emergence of level crossing. The transition frequencies of the $R$-branch, $b$-type transition series with $J''=17$, $K_a''=K_a'+1$, and $K_a''+K_c''=J''$ 
  are plotted from $K_a''=12-17$ with respect to the offset from the center frequency $f_c$ in each panel. 
  Sticks in blue represent the lines with normal $b$-type selection rule, where $\Delta K_c=-1$.
  Sticks in brown represent the lines with $c$-type selection rule due to level-crossing, where $\Delta K_c = 0$. 
  The height of the sticks is proportional to the relative intensity of these components. 
  The separation between the $b$-type components and $c$-type components at $K_a''=14$ is the largest.
   \label{fig:level-crossing}}
\end{figure}

In our 2-propanimine spectrum, 
%Despite this restriction of selection rules by the dipole moment, we did predict and observe $c$-type transitions in other cases when the $\mu_c$ dipole moment was set to 0. 
these mis-labeled $c$-type transitions appear when $K_c$ is sufficiently small compared to $K_a$, 
and they only accompany $b$-type transitions and not $a$-type transitions.
%In these cases, the $c$-type lines arise due to mixing of asymmetric rotor wavefunctions and level crossing, 
%which is a phenomenon observed and discussed in similar two-top systems, such as in dimethyl ether \citep{Endres2009AA}, and ethyl methyl ether \citep{Fuchs2003ApJS}. 
Figure~\ref{fig:level-crossing} demonstrates the emergence of level crossing for the $b$-type, $R$-branch, $J=17\leftarrow 16$ lines. 
When $K_a'' \leq 13$, rotational quantum numbers with normal $b$-type selection rule ($\Delta K_a=1$, $\Delta K_c=-1$) 
correctly label the five torsional components with correct relative intensity.
At $K_a'' = 14$, the tunneling splitting pattern abruptly alters into 9 components, in which five components remain the $b$-type selection rule, 
and four new components appear with $c$-type selection rule $\Delta K_c=0$.
The separation between the $b$-type and $c$-type components is approximately 500~MHz, one order of magnitude larger than the span of the regular torsional splitting, which rarely exceeds 50~MHz. 
These extra $c$-type components arise from the mixing of nearly degenerate asymmetric rotor wavefunctions,
and intensities are redistributed between them and their $b$-type counterparts.
As $K_a''$ continues to increase, the tunneling splitting pattern resumes to five components,
and the $b$-type selection rule is only maintained in the $(\sigma_1,\sigma_2)=(0,0)$ state, 
as it corresponds to the torsional sublevel of the lowest energy that is not affected by the mislabeling. 
%These irregular selection rules only affect $b$-type lines, and they occur only when $K_c$ becomes sufficiently small. 
%In these level-crossing cases, the center frequency $f_c$ of all torsional splitting components starts to deviate from the pure asymmetric top model.
%In Figure~\ref{fig:split-pattern}, such trend begins to appear at the lowest $J$ value.

\subsection{Fit statistics}\label{app:fit}

A detailed error analysis shows that in the XIAM model, 
97.89~\% of the fitted transitions have frequency deviations within $3\sigma$. 
575 transitions (out of 27,259 transitions) have frequency deviations larger than $3\sigma$, 
in which 152 deviations are $>5\sigma$. 
The maximum frequency deviation is $22\sigma$, which is an indication of the insufficiency of the model. 
The transitions with the largest deviations are those with high $K_a$ values. 
This result is expected because XIAM has a limited number of high order correction terms 
to treat a data set with such high $J$ and $K_a$ values. 
In the \mbox{ERHAM} fit, 99.91~\% of the fitted transitions have frequency deviations within $3\sigma$, 
and the deviations of the rest 0.09~\% transitions are all within $5\sigma$.
These statistics indicate that the ERHAM fit has reached experimental accuracy.

\section{Input and output files for XIAM and ERHAM}\label{app:c}

The input and output files for XIAM and ERHAM fit can be found 
on Zenodo (DOI: 10.5281/zenodo.7541890) as \verb|TableC1.xi|, 
\verb|TableC1.xo|, \verb|TableC2.in|, and \verb|TableC2.out|.

% Don't change these lines
\bsp	% typesetting comment
\label{lastpage}
\end{document}